\documentclass[12pt,preprint,natbib,iop,numberedappendix,appendixfloats]{emulateapj} %Use this for all other purposes (emulateapj)
\usepackage{graphicx,epsfig,amsmath} %Comment out hyperref for submission

\lefthead{van der Wel et al.}
\righthead{Size-Mass Relation from CANDELS/3D-HST}
\slugcomment{Accepted for publication in ApJ}
\begin{document}

\newcommand{\gala}{{\tt GALAPAGOS}~}
\newcommand{\gf}{{\tt GALFIT}~}
\newcommand{\mh}{H_{\rm{F160W}}}
\newcommand{\kms}{\>{\rm km}\,{\rm s}^{-1}}
\newcommand{\reff}{R_{\rm{eff}}}
\newcommand{\msol}{M_{\odot}}
\newcommand{\msola}{10^{11}~M_{\odot}}
\newcommand{\msolb}{10^{10}~M_{\odot}}
\newcommand{\msolc}{10^{9}~M_{\odot}}

\title{3D-HST+CANDELS: THE EVOLUTION OF THE GALAXY SIZE-MASS
  DISTRIBUTION SINCE $z=3$}

\author{A.~van der Wel\altaffilmark{1}}
\author{M.~Franx\altaffilmark{2}}
\author{P.G.~van Dokkum\altaffilmark{3}}
\author{R.E.~Skelton\altaffilmark{4}}
\author{I.G.~Momcheva\altaffilmark{3}}
\author{K.E.~Whitaker\altaffilmark{5}}
\author{G.B.~Brammer\altaffilmark{6}}
\author{E.F.~Bell\altaffilmark{7}}
\author{H.-W.~Rix\altaffilmark{1}}
\author{S.~Wuyts\altaffilmark{8}}
\author{H.C.~Ferguson\altaffilmark{6}}
\author{B.P.~Holden\altaffilmark{9}}
\author{G.~Barro\altaffilmark{9}}
\author{A.M.~Koekemoer\altaffilmark{6}}
\author{Yu-Yen Chang\altaffilmark{1}}
\author{E.J.~McGrath\altaffilmark{10}}
\author{B.~H\"aussler\altaffilmark{11} \altaffilmark{12}}
\author{A.~Dekel\altaffilmark{13}}
\author{P.~Behroozi\altaffilmark{6}}
\author{M.~Fumagalli\altaffilmark{2}}
\author{J.~Leja\altaffilmark{3}}
\author{B.F.~Lundgren\altaffilmark{14}}
\author{M.V.~Maseda\altaffilmark{1}}
\author{E.J.~Nelson\altaffilmark{3}}
\author{D.A.~Wake\altaffilmark{14} \altaffilmark{15}}
\author{S.G.~Patel\altaffilmark{16}}
\author{I.~Labb\'e\altaffilmark{2}}
\author{S.M.~Faber\altaffilmark{9}}
\author{N.A.~Grogin\altaffilmark{6}}
\author{D.D.~Kocevski\altaffilmark{17}}

\altaffiltext{1}{Max-Planck Institut f\"ur Astronomie, K\"onigstuhl
  17, D-69117, Heidelberg, Germany; e-mail:vdwel@mpia.de}

\altaffiltext{2}{Leiden Observatory, Leiden University, P.O.Box 9513,
  NL-2300 AA Leiden, The Netherlands}

\altaffiltext{3}{Department of Astronomy, Yale University, New Haven,
  CT 06511, USA}

\altaffiltext{4}{South African Astronomical Observatory, PO Box 9,
  Observatory, 7935, South Africa}

\altaffiltext{5}{Astrophysics Science Division, Goddard Space Center,
  Greenbelt, MD 20771, USA}

\altaffiltext{6}{Space Telescope Science Institute, 3700 San Martin
  Drive, Baltimore, MD 21218, USA}

\altaffiltext{7}{Department of Astronomy, University of Michigan, 500
  Church Street, Ann Arbor, MI 48109, USA}

\altaffiltext{8}{Max-Planck Institut f\"ur Extraterrestrische Physik,
  Giessenbachstrasse, D-85748 Garching, Germany}

\altaffiltext{9}{University of California Observatories/Lick
  Observatory, University of California, Santa Cruz, CA 95064, USA}

\altaffiltext{10}{Department of Physics and Astronomy, Colby College,
  Waterville, ME 0490, USA}

\altaffiltext{11}{Physics Department, University of Oxford, Denys
  Wilkinson Building, Keble Road, Oxford OX1 3RH, UK}

\altaffiltext{12}{Centre for Astrophysics, Science \& Technology
  Research Institute, University of Hertfordshire, Hatfield, Herts,
  AL10 9AB, UK}

\altaffiltext{13}{Racah Institute of Physics, The Hebrew University,
  Jerusalem 91904, Israel}

\altaffiltext{14}{Department of Astronomy, University of Wisconsin,
  Madison, WI 53706, USA}

\altaffiltext{15}{The Department of Physical Sciences, The Open
  University, Milton Keynes, MK7 6AA, UK}

\altaffiltext{16}{Observatories of the Carnegie Institution of
  Washington, Pasadena, CA 91101, USA}

\altaffiltext{17}{Department of Physics and Astronomy, University of
  Kentucky, Lexington, KY 40506, USA}

\begin{abstract}
  Spectroscopic$+$photometric redshifts, stellar mass estimates, and
  rest-frame colors from the 3D-HST survey are combined with
  structural parameter measurements from CANDELS imaging to determine
  the galaxy size-mass distribution over the redshift range $0<z<3$.
  Separating early- and late-type galaxies on the basis of
  star-formation activity, we confirm that early-type galaxies are on
  average smaller than late-type galaxies at all redshifts, and we
  find a significantly different rate of average size evolution at
  fixed galaxy mass, with fast evolution for the early-type
  population, $\reff\propto (1+z)^{-1.48}$, and moderate evolution for
  the late-type population, $\reff\propto (1+z)^{-0.75}$.  The large
  sample size and dynamic range in both galaxy mass and redshift, in
  combination with the high fidelity of our measurements due to the
  extensive use of spectroscopic data, not only fortify previous
  results, but also enable us to probe beyond simple average galaxy
  size measurements.  At all redshifts the slope of the size-mass
  relation is shallow, $\reff\propto M_*^{0.22}$, for late-type
  galaxies with stellar mass $>3\times \msolc$, and steep,
  $\reff\propto M_*^{0.75}$, for early-type galaxies with stellar mass
  $>2\times \msolb$.  The intrinsic scatter is $\lesssim$0.2 dex for
  all galaxy types and redshifts.  For late-type galaxies, the
  logarithmic size distribution is not symmetric but is skewed toward
  small sizes: at all redshifts and masses a tail of small late-type
  galaxies exists that overlaps in size with the early-type galaxy
  population.  The number density of massive ($\sim \msola$), compact
  ($\reff < 2$kpc) early-type galaxies increases from $z=3$ to
  $z=1.5-2$ and then strongly decreases at later cosmic times.
\end{abstract}

\section{Introduction}\label{sec:intro}

The size distribution of the stellar bodies of galaxies, and its
evolution with cosmic time, provides important clues about the
assembly history of galaxies and the relationship with their dark
matter halos.  The two main classes of galaxies, early and late types,
show very different dependencies between size and stellar mass
\citep{shen03}.  The weak dependence between size and mass for
late-type galaxies implies that the high-mass late types, on average,
have higher surface mass densities than low-mass late types.  In
contrast, early types show a more complex relationship between stellar
mass and density, with the density peaking for systems with stellar
masses around $M_*\sim 4\times 10^{10}~\msol$ and decreasing toward
both lower and higher masses, as reflected in the classical
\citet{kormendy77} relation.  This fundamental difference does not
depend on whether classification of early and late types is based on
star formation activity, bulge dominance (S\'ersic index), or visual
inspection, and it implies that the two types have very different
evolutionary and assembly histories.

In this paper we present the evolution of the size-mass distribution
up to $z=3$ on the basis of spectroscopy and multiwavelength
photometry from the 3D-HST survey \citep{brammer12a} and
\emph{HST}/WFC3 imaging from CANDELS \citep{grogin11,koekemoer11}.
Angular galaxy sizes are measured from the CANDELS imaging as
described by \citet{vanderwel12} and the \emph{HST}/WFC3 grism
observations from 3D-HST provide spectroscopic confirmation and
redshifts for a large fraction of the sample, considerably
strengthening -- with respect to previous studies -- the fidelity of
estimates for stellar masses and rest-frame photometric properties.

So far, most of the previous studies have focused on the evolution of
average galaxy sizes of the high-mass end of the distribution
($\gtrsim 5\times \msolb$).  Enabled by both the improved data quality
and a fivefold increase in sample size, we can now, for the first
time, describe the size distribution of galaxies across redshift.

\subsection{Size Evolution of Late-type Galaxies}\label{sec:introlate}

Tracing the evolution of the size distribution with redshift allows us
to test the most basic elements in our theory of galaxy formation. The
zeroth-order expectation is that disk scale lengths evolve fast,
approximately as the inverse of the Hubble parameter \citep{mo98}, and
early and recent work on the average sizes of Lyman break galaxies
(LBGs) at high redshifts ($z\sim 2-6$) roughly agree with this
expectation for a $\Lambda$CDM cosmology: \citet{giavalisco96},
\citet{ferguson04}, \citet{oesch10}, and \citet{mosleh12} all find
rapid size evolution with redshift: $\reff\propto (1+z)^{\beta=-1.1}$.

In contrast, the average size at a given stellar mass of the
population of disk-dominated galaxies evolves slowly at late times
($z\lesssim 1$) and has been reported to evolve slowly as measured at
fixed galaxy mass ($\beta = -0.2$) or not at all \citep{lilly98,
  ravindranath04, barden05}.  The implication would be that the
evolution of the disk galaxy population is decoupled from the
evolution of the dark matter halo population.  One fundamental
difference between the results of LBGs and lower-redshift disk
galaxies is the rest-frame wavelength at which the sizes are measured:
the rest-frame UV light seen for LBGs originates from young stars that
may be, and are generally expected to be, distributed differently than
bulk of the baryonic and stellar mass, not to mention the consequences
of extinction.

The advent of ground-based near-infrared imaging surveys helped to
bridge the $z<1$ and $z>2$ regimes by enabling size measurements in a
consistent manner at a fixed rest-frame wavelength.  Early results
suggested slow evolution for late-type galaxies up to $z\sim 3$
\citep{trujillo06a}, but the uncertainties at $z>1$ were such that
evolution in that regime was not strongly constrained.  Later
ground-based work pointed at faster evolution at a fixed galaxy mass:
\citet{franx08} found $\beta=-0.6$ and \citet{williams10} found
$\beta=-0.9$, but precise constraints at $z>1.5$ remained elusive and
the apparent tension between the $z\lesssim 1$ work and the
near-infrared at $z\sim 1.5$ unaddressed.

Several \emph{HST}/NICMOS-based studies of the morphology and
structure of massive $z\sim 2$ galaxies in the rest-frame optical
eventually led to mostly converged results, with $\beta\sim -0.8$ for
massive ($\sim \msola$), star-forming galaxies from $z\sim 2.5$ to the
present \citep{toft07,buitrago08,kriek09b}.  So far it has remained
unclear as to whether the difference with the significantly faster
evolution for LBG galaxies ($\beta\sim -1.1$) is caused by
morphological $K$ corrections, the difference in mass (the typical LBG
has $M_*\sim \msolb$), or physical changes with redshift.  In
addition, the difference with the previously mentioned studies at
$z<1$ \citep{lilly98, ravindranath04, barden05} remains unexplained.

Improving the measurement of $\beta$ and its mass dependence is
crucial in order take the next step toward understanding disk galaxy
formation.  In this paper we will address these issues and describe
the full size-mass distribution of high-redshift galaxies over a broad
range in galaxy mass and redshift.  We will

\begin{itemize}
\item{measure the evolution of the slope of the size-mass relation;}
\item{present the size distribution as a function of stellar mass and
    redshift;}
\item{provide a consistent comparison with UV-selected, high-redshift
    samples.}
\end{itemize}

\subsection{Size Evolution of Early-type Galaxies}\label{sec:introearly}

Over the past five years, more attention has been bestowed on the size
evolution of early-type galaxies than on the size evolution of
late-type, star-forming galaxies.  Interest in the topic was initiated
by reports that $z\sim 1.5$ early-type galaxies have remarkably small
sizes in \emph{HST}-based rest-frame UV imaging
\citep{daddi05,trujillo07} and ground-based near-infrared imaging
\citep{trujillo06b}.  NICMOS imaging presented by \citet{zirm07},
\citet{toft07}, \citet{stockton08}, and \citet{mcgrath08} provided
space-based, rest-frame optical size measurements that strengthened
the evidence for rapid size evolution ($\beta= -1$ or faster) as
measured at a fixed galaxy mass ($\sim\msola$).  This notion became
firmly established through larger samples \citep{buitrago08} and the
first spectroscopic samples \citep{vandokkum08, damjanov11}.

Concerns regarding gross overestimates of the stellar mass content of
the compact early-type galaxies were alleviated by dynamical mass
estimates of such galaxies at $z\gtrsim 1$
\citep{vanderwel08,cimatti08,newman10,vandesande11,toft12,vandesande13,bezanson13,belli14},
and the analysis by \citet{szomoru10} of ultra-deep imaging of a
single compact galaxy has demonstrated the absence of low-surface
brightness wings that could have been missed by shallower imaging.
Increases in sample size and dynamic range in stellar mass have
constrained the average size evolution of early-type galaxies with
stellar masses $>5\times \msolb$ to $\beta\sim -1.3$ up to $z=2.5$,
with no evidence for a change in the slope of the relation over this
mass range \citep{newman12}.  The steepness of the relation combined
with the non-negligible scatter accommodates observations that
early-type galaxies display a large range in size at $z>1$
\citep[e.g.,][]{mancini10,saracco11}.

While the observational results have largely converged, the
interpretation is still debated.  Some authors have considered the
average increase in size over time as being due to the addition of
new, larger early-type galaxies.  While some argue that this cannot
reproduce the observations \citep{vanderwel09a}, others argue that a
population of compact early-type galaxies (with sizes $\reff\lesssim
2$kpc) exists within present-day clusters, with a number density
comparable to that of higher-redshift early-type galaxies
\citep{valentinuzzi10,poggianti13}; tension with the absence of such
galaxies in the Sloan Digital Sky Survey (SDSS) remains
\citep{trujillo09, taylor10}.  The crucial observational test is to
trace the evolution of the number density of early-type galaxies as a
function of their size. \citet{cassata11}, \citet{szomoru12}, and
\citet{newman12} show strong evolution in the number density of small
galaxies at $0<z<2.5$, while \citet{carollo13} claim no evolution at
$0<z<1$.  Our use of 5 fields addresses the issue of field-to-field
variations that may affect the aforementioned studies based on smaller
samples, and it extends the dynamic redshift range of the Carollo et
al.~sample.

The leading explanation for the size growth of individual galaxies is
accretion and tidal disruption of satellite galaxies that gradually
build up the outer parts.  For this process, the change in size is
large compared with the increase in mass: $\Delta \reff \propto \Delta
M_*^2$ \citep[e.g.,][]{bezanson09,hopkins09}.  This analytical
prediction based on conservation of binding energy has been tested
through numerical simulations \citep[e.g.,][]{naab09,oser12,
  bedorf13}.  The analytically predicted and simulated evolution in
the increased surface mass density at large radii is, in fact,
observed \citep{vandokkum10}; in addition, the central stellar density
shows little evolution \citep{bezanson09,hopkins09,vandokkum10}, which
is also consistent with a minor merger scenario.  In other words, the
observations show that there is no need, and little room, to
physically expand a galaxy by displacing large numbers of stars to
large radii through rapid changes in the central potential, as
suggested by \citet{fan10}.  A possible challenge to the minor merger
scenario is posed by the lack of strong evolution in the slope of the
mass density profile seen in lensing galaxies \citep{sonnenfeld13}.

Until recently, the size evolution of late- and early-type galaxies
was usually discussed separately and treated as different topics.
However, in order to understand the joint evolution of these classes,
one has to take into account the continuous transition of late-type to
early-type galaxies seen in particular in the stellar mass range of
$\msolb$ to $\msola$
\citep[e.g.,][]{bell04,faber07,brown07,ilbert10,brammer11,buitrago13,muzzin13}.
The star-forming progenitors of the small early-type galaxies are now
looked for and plausibly identified \citep[e.g.,][]{whitaker11,
  barro13, barro14, toft14}, but the evolutionary path of the
transitioning galaxies has not been fully mapped out.

In this paper, regarding the evolution of early-type galaxies, we will

\begin{itemize}
\item{present the distribution of sizes as a function of stellar mass
    and redshift, jointly with those of late-type galaxies; and}
\item{show the evolution of the number density of early-type galaxies
    as a function of size.}
\end{itemize}

After describing the data and sample selection in Section
\ref{sec:data} we present and analyze size distributions as a function
of redshift and galaxy mass in Section \ref{sec:distr}.  We compare
our results with previous studies in Section \ref{sec:comparison} and
then discuss the implications of our findings in Section
\ref{sec:discussion}.  We assume the cosmological parameters
$(\Omega_{\rm{M}},~\Omega_{\Lambda},~h) =
(0.27,~0.73,~0.71)$. Finally, we use AB magnitudes and the
\citet{chabrier03} stellar initial mass function.

\begin{figure*}[t]
\epsscale{1} 
%\epsscale{0.5} 
\plotone{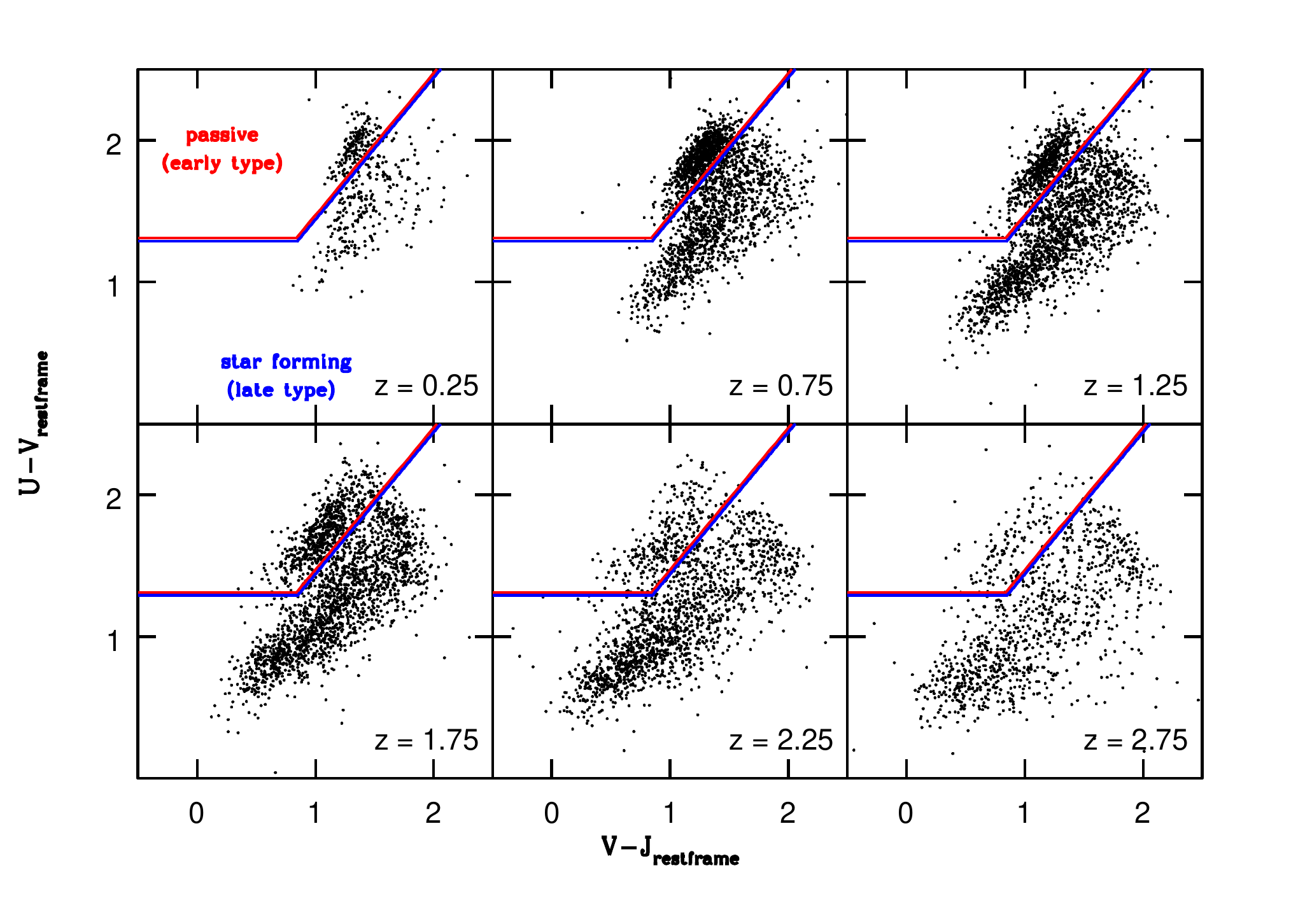}
\caption{Rest-frame $U-V$ vs.~$V-J$ color distribution for six
  redshift bins (each 0.5 wide).  The two distinct classes of
  quiescent and star-forming galaxies are separated by the indicated
  selection criteria to define our early- and late-type galaxy
  samples.}
\label{uvj}
\end{figure*}

\begin{figure*}[t]
\epsscale{1} 
%\epsscale{0.5} 
%\includegraphics[angle=-90,scale=0.6]{M_mag_mosaic.pdf}
\plotone{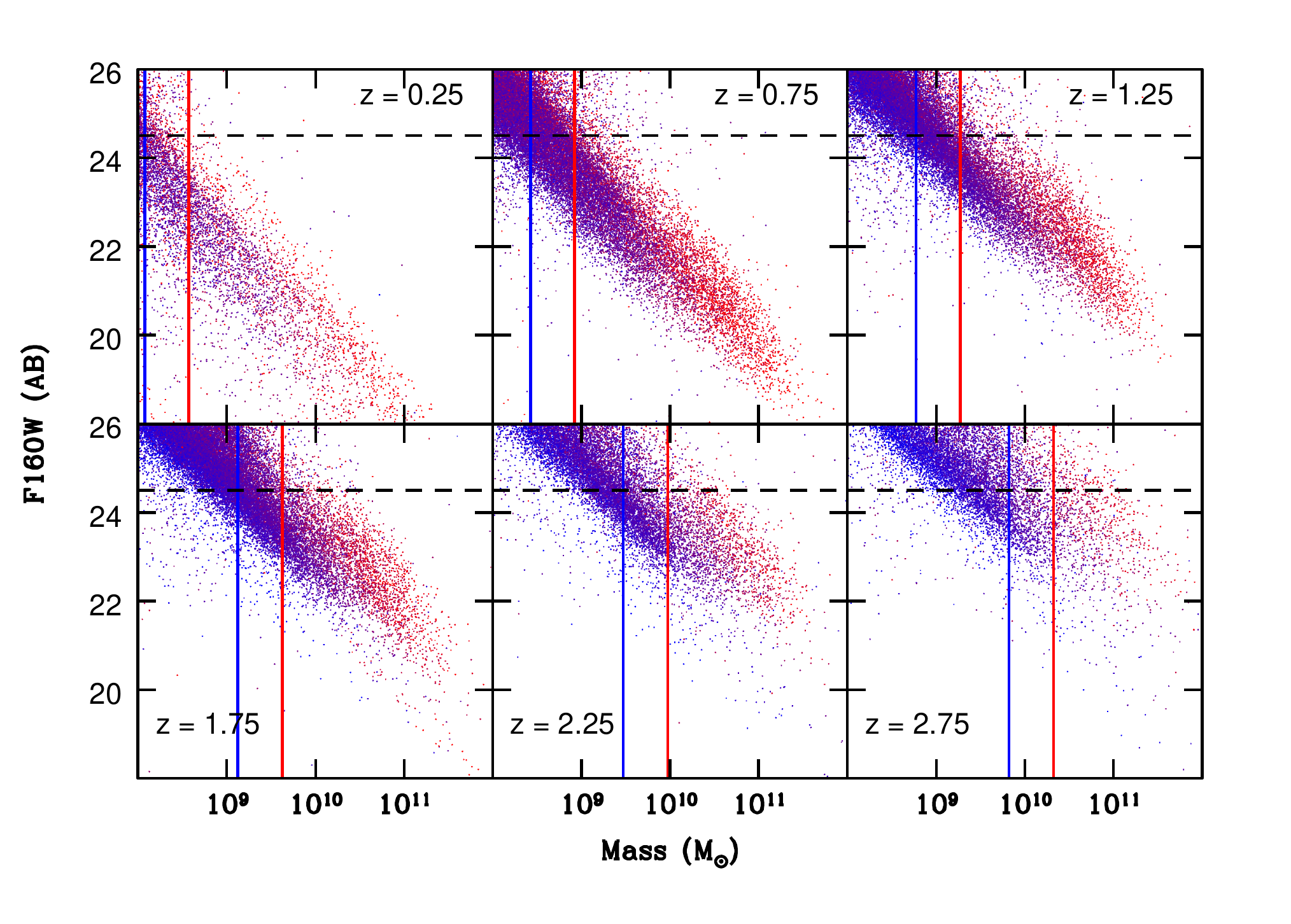}
\caption{Observed $\mh$ magnitude vs.~stellar mass in six redshift
  bins.  The color coding represents the rest-frame $U-V$ colors,
  ranging from $U-V=0$ (blue) to $U-V=2$ (red).  The horizontal dashed
  lines indicate the limit ($\mh=24.5$) down to which we can determine
  sizes with good fidelity.  The vertical lines illustrate the
  resulting mass completeness limits for blue ($U-V=0.5$) and red
  ($U-V=2.0$) galaxies, respectively.  See Section \ref{sec:sample}
  for further details.}
\label{mmag}
\end{figure*}

\begin{figure*}[t]
\epsscale{1} 
%\epsscale{0.5} 
\plotone{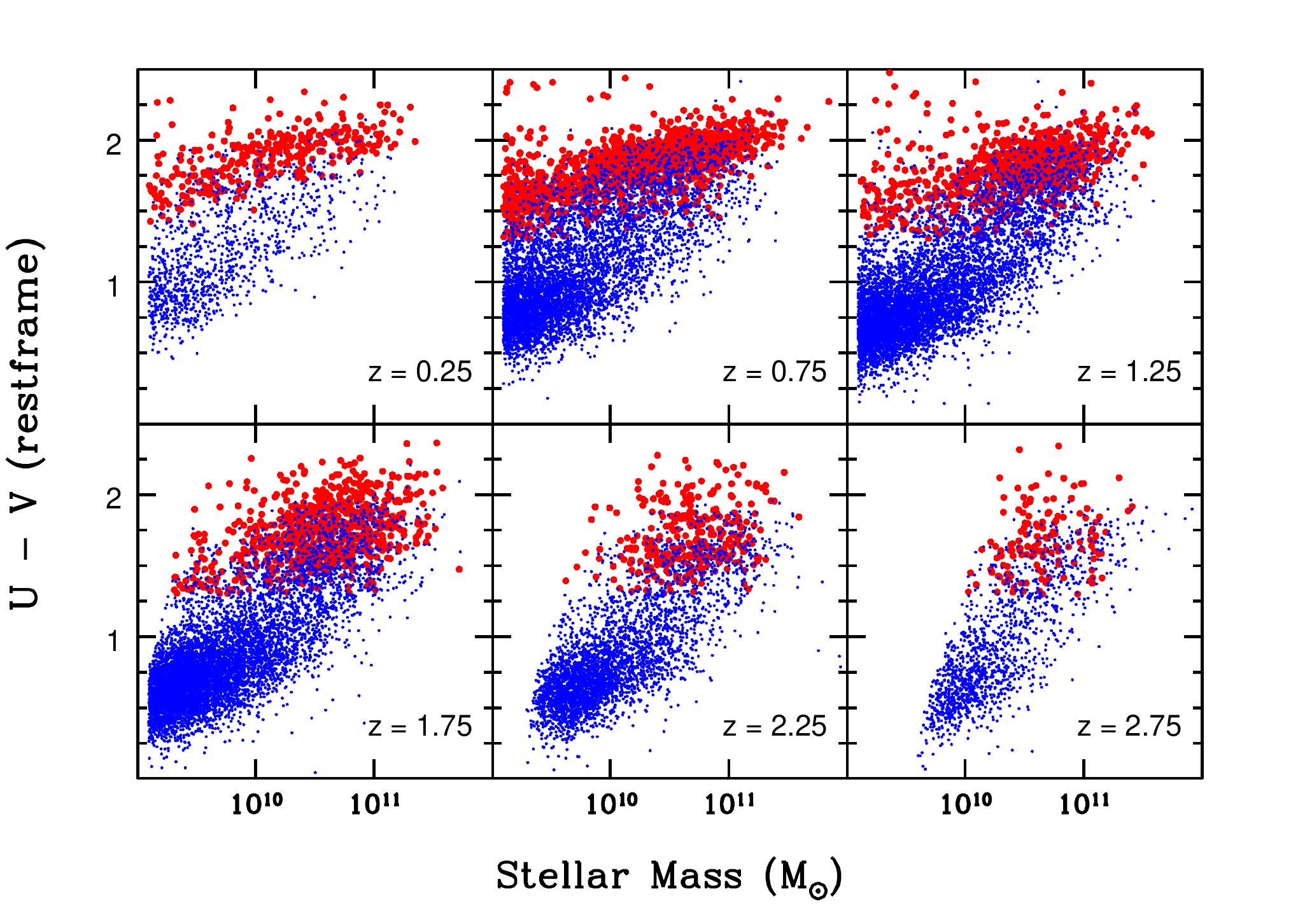}
\caption{Rest-frame $U-V$ color vs. stellar mass in six redshift bins.
  Early-type galaxies, defined as illustrated in Figure \ref{uvj}, are
  shown in red and late types in blue.  A clearly defined red sequence
  is seen up to $z=3$, with an increased prevalence of dusty late-type
  galaxies toward higher redshifts.}
\label{muv}
\end{figure*}

\begin{figure}[t]
\epsscale{1.2} 
%\epsscale{0.5} 
\plotone{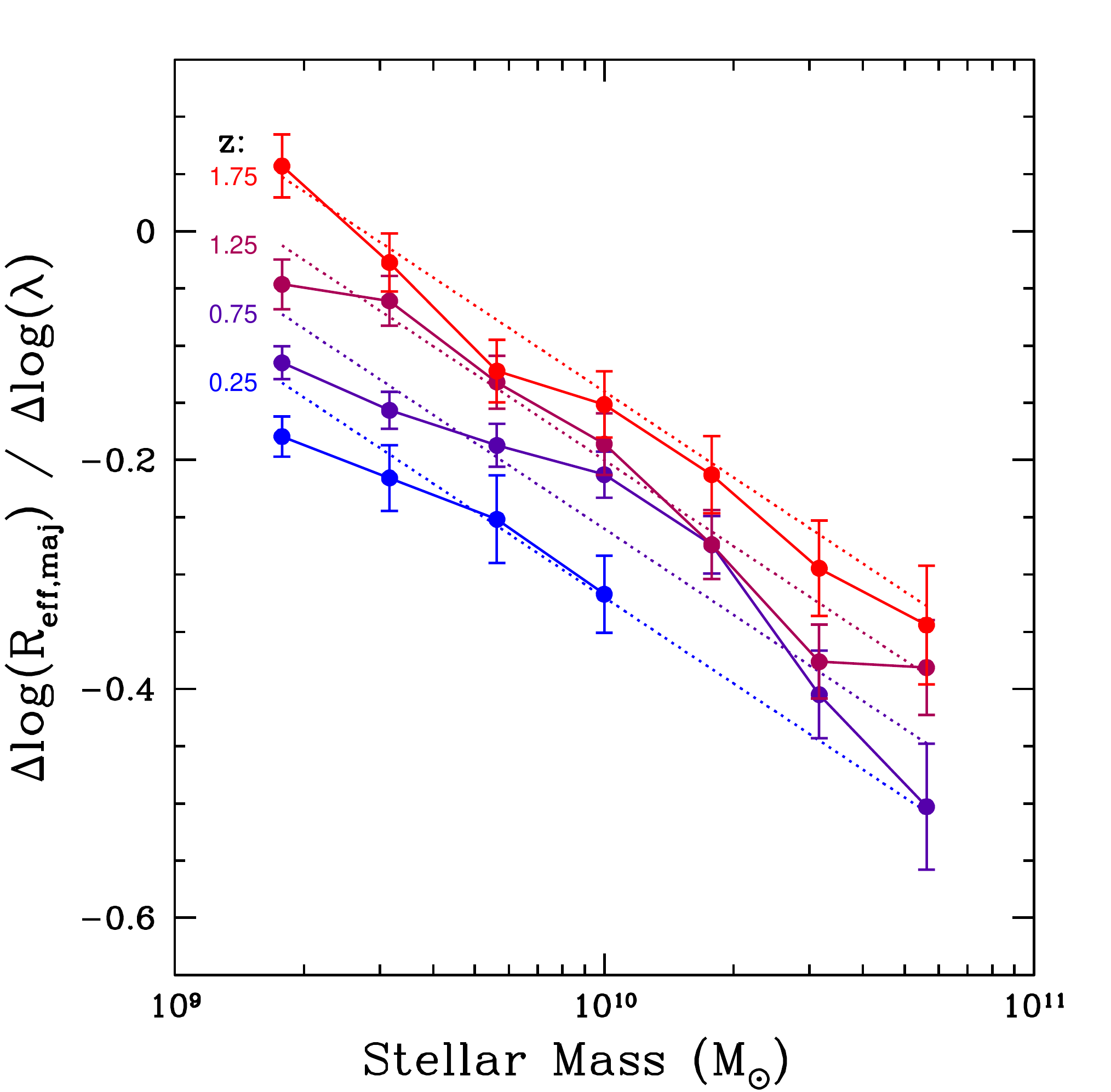}
\caption{Wavelength dependence of $\reff$ in bins of stellar mass and
  redshift; the latter is indicated by the color coding.  Late-type
  galaxies, as defined in Figure \ref{uvj}, with robust size
  measurements from ACS/F814W, WFC3/F125W and WFC3/F160W imaging are
  included (see text for details).  Generally, sizes are smaller at
  longer wavelengths, that is, late-type galaxies are bluer in the
  outer parts.  Moreover, this gradient is stronger for more massive
  galaxies at all redshifts, and the gradient decreases with redshift,
  at the same rate for all masses.  The dotted lines represent the
  parameterization given in Equation~(\ref{eq:grad}) that we use to
  correct our size measurements of late-type galaxies.}
\label{mgrad}
\end{figure}

\begin{figure*}[t]
\epsscale{1} 
%\epsscale{0.5} 
\plotone{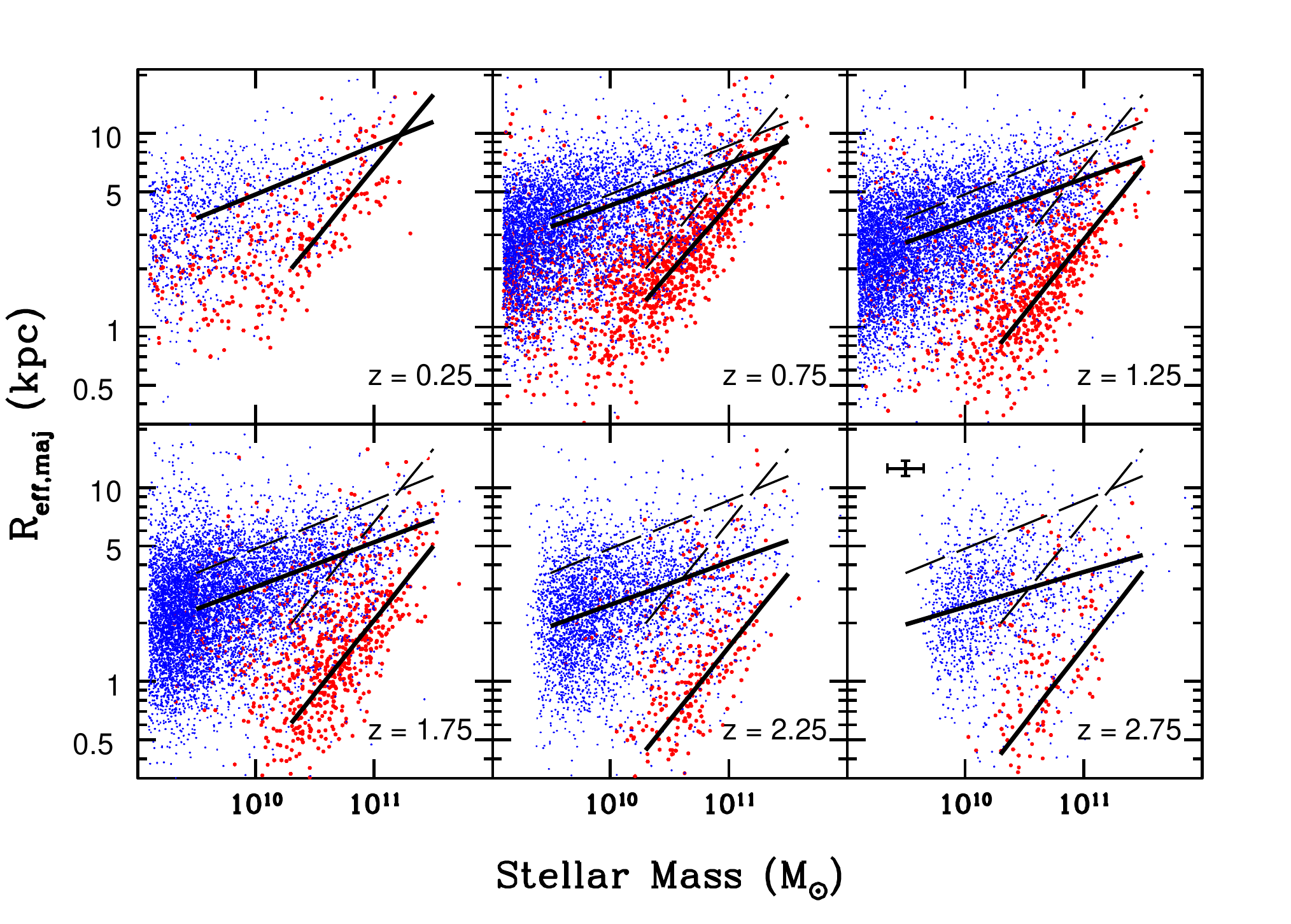}
\caption{Size-stellar mass distribution of late- and early-type
  galaxies (same symbols as in Figure \ref{mmag}).  A typical
  $1\sigma$~error bar for individual objects in the higher-redshift
  bins is shown in the bottom-right panel.  The lines indicate model
  fits to the early- and late-type galaxies as described in Section
  \ref{sec:fitsize}.  The dashed lines, which are identical in each
  panel, represent the model fits to the galaxies at redshifts
  $0<z<0.5$.  The solid lines represent fits to the higher-redshift
  samples. The mass ranges used in the fits are indicated by the
  extent of the lines in the horizontal direction.  Strong evolution
  in the intercept of the size-mass relation is seen for early-type
  galaxies and moderate evolution is seen for the late-type galaxies
  (also see Figure \ref{parevol}).  There is no significant evidence
  for evolution in the slope (also see Figure \ref{parevol}). The
  parameters of the fits shown here are given in Table
  \ref{tab:fitsize}.}
\label{mr}
\end{figure*}

\begin{figure*}[t]
%\epsscale{1.2} 
\epsscale{0.38} 
\plotone{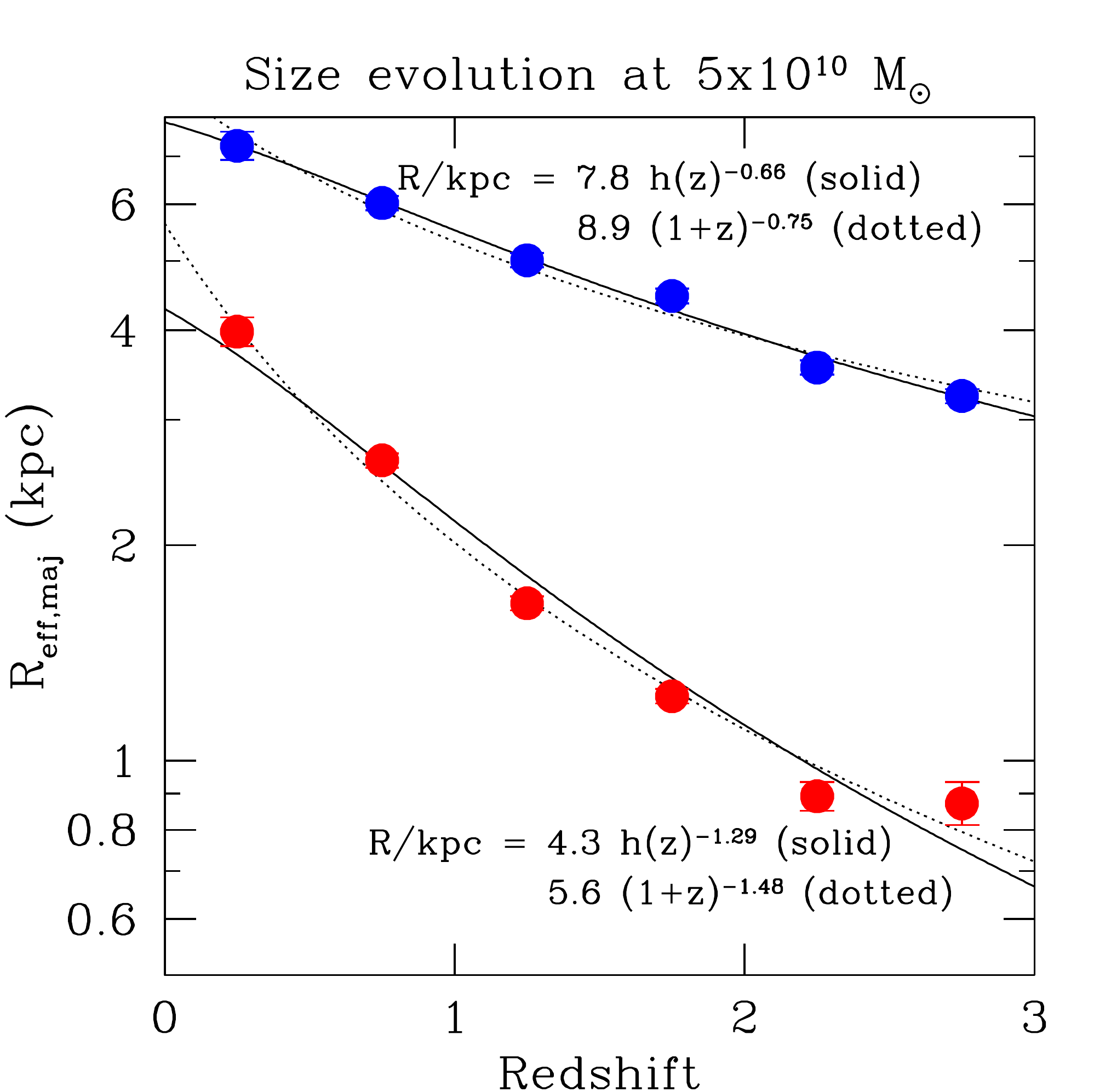}
\plotone{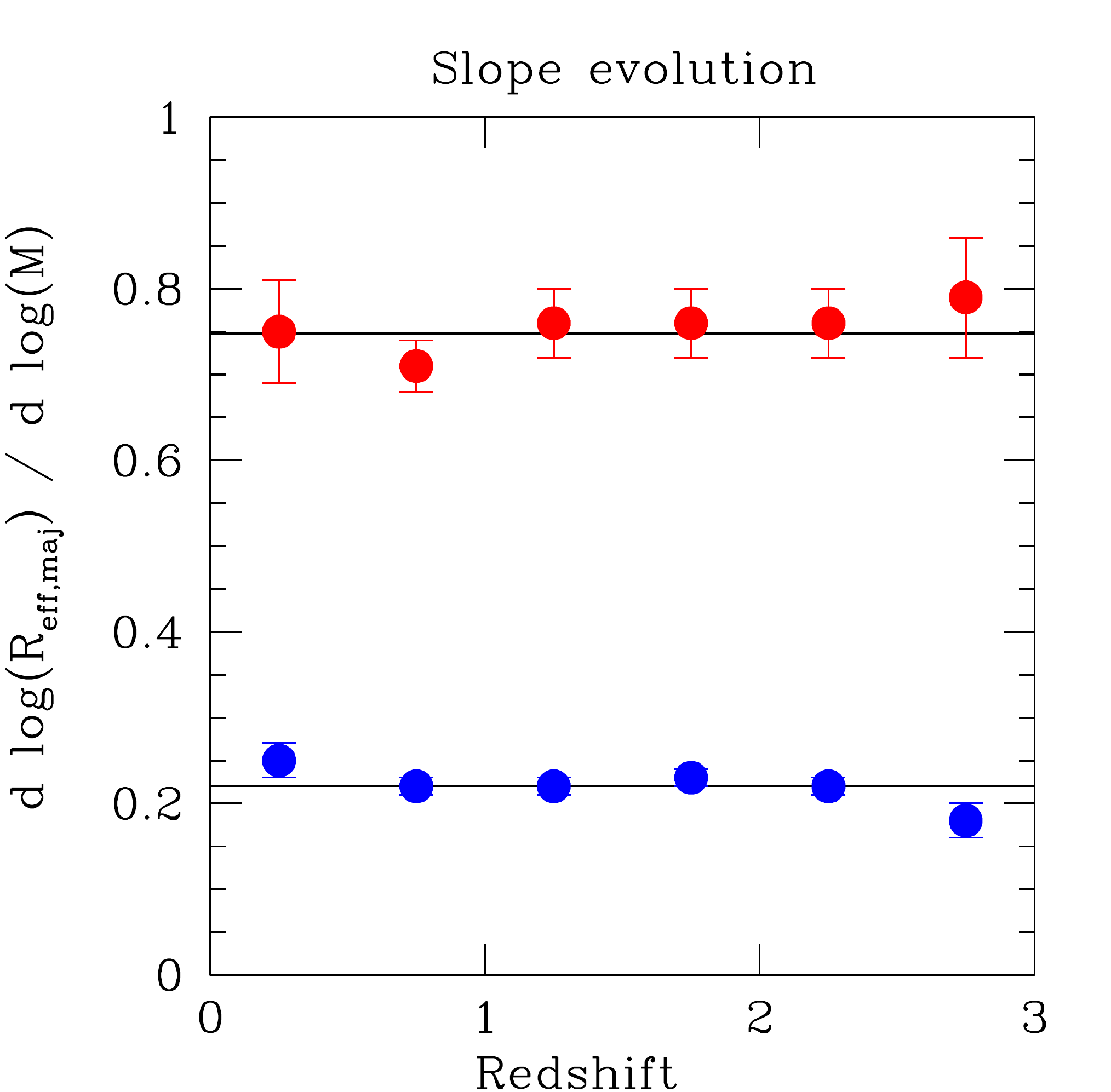}
\plotone{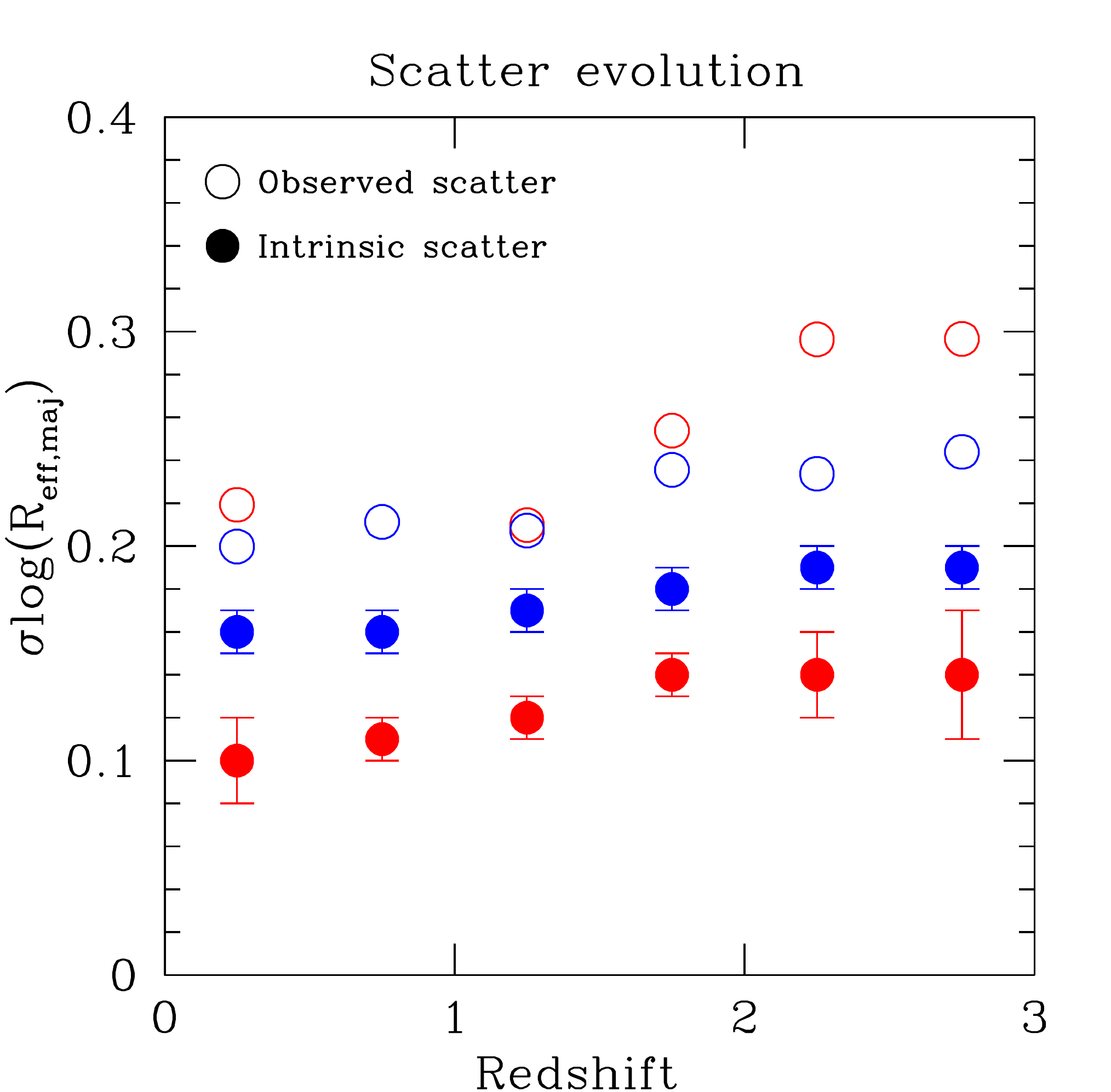}
\caption{Parameterized redshift evolution of the size-mass relation,
  from the power law model fits shown in Figure \ref{mr}. The
  left-hand panel shows the evolution of the intercept, or the size
  evolution at fixed stellar mass of $5\times 10^{10}~\msol$.  Strong
  evolution is seen for high-mass early-type galaxies and moderate
  evolution is seen for low-mass early types and for late-type
  galaxies.  The middle and right-hand panels show the evolution of
  the slope and intrinsic (model) scatter of the size-mass relation,
  either with little or no evidence for changes with redshift.  The
  open symbols represent the observed scatter: these measurements do
  not take measurement uncertainties and contamination into account.
  The fitting parameters shown in this figure are given in Table
  \ref{tab:fitsize}.}
\label{parevol}
\end{figure*}

\begin{figure}[t]
\epsscale{1.2} 
\plotone{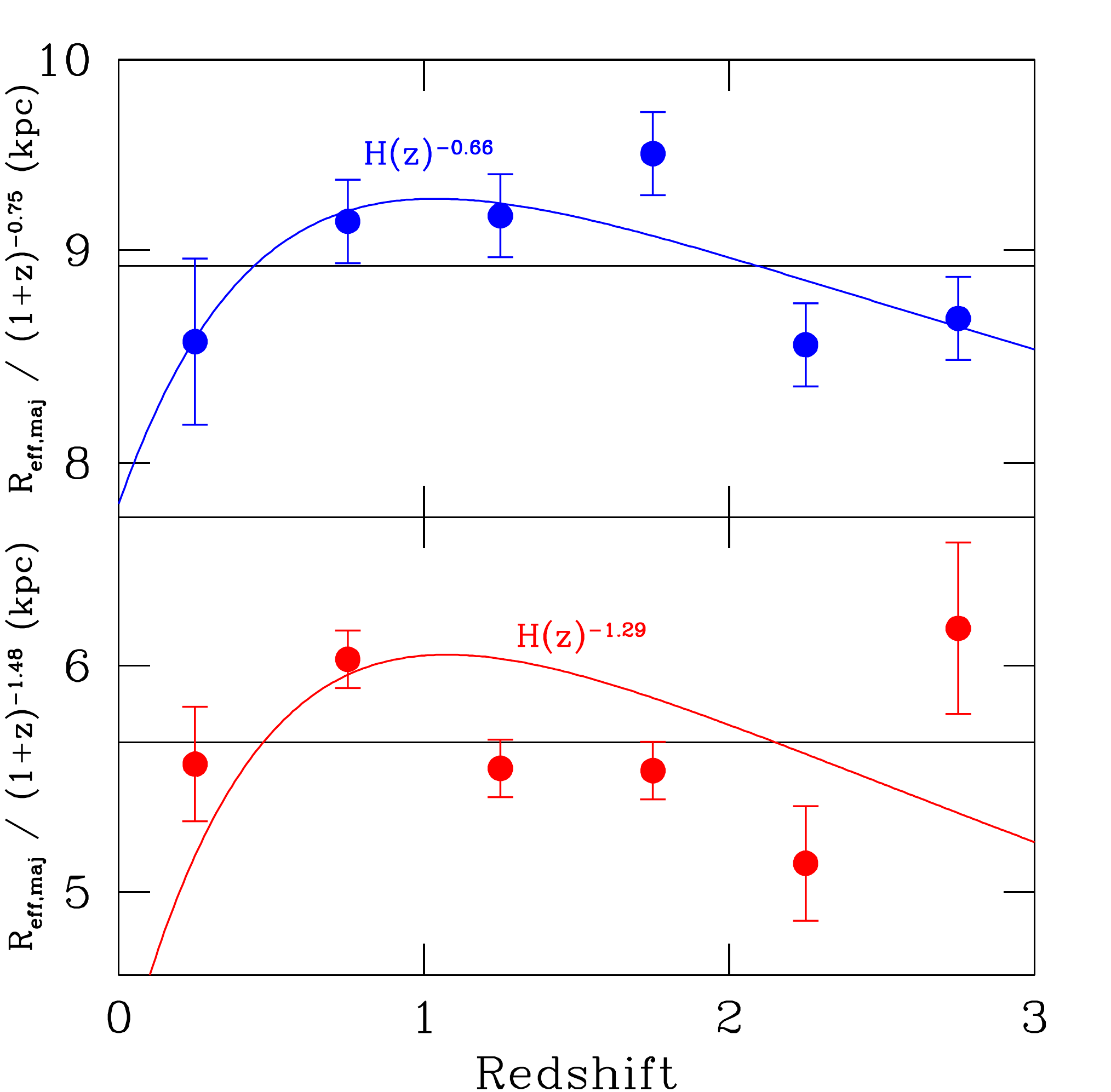}
\caption{Evolution-corrected average sizes at $M_*=5\times\msolb$ for
  late-type galaxies (top panel, in blue) and early-type galaxies
  (bottom panel, in red).  The values shown here are the values shown
  in the left-hand panel of Figure \ref{parevol}, divided by
  $(1+z)^{\beta_z}$ as indicated on the y-axis.  The residuals from
  the best-fitting $(1+z)^{\beta_z}$ law indicate that parameterizing
  the evolution as a function of the Hubble parameter ($\reff\propto
  h(z)^{\beta_H}$) may provide a more accurate description of the
  late-type galaxies.  See Section \ref{sec:medsize} for further
  discussion.}
\label{zr_res}
\end{figure}

\begin{figure*}[t]
\epsscale{1.2} 
%\epsscale{0.5} 
\plotone{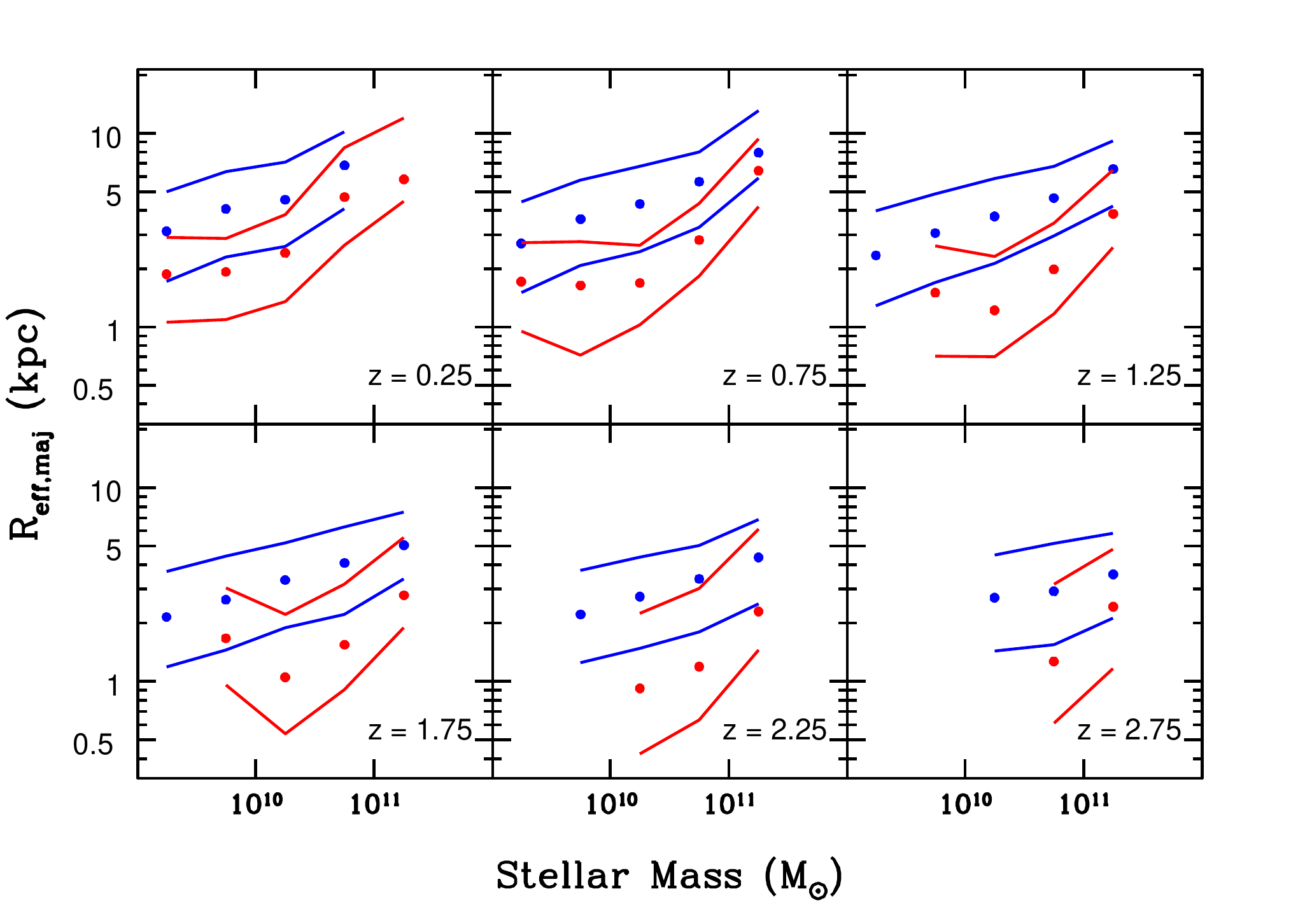}
\caption{Median (points) and 16th and 84th percentiles (lines) of the
  size-stellar mass distributions shown in Figure \ref{mr}.  The
  scatter in $\reff$ does not strongly depend on galaxy mass.
  Deviations from the power law form of the size-mass relation are
  clear for massive late-type galaxies and for low-mass early-type
  galaxies.  Note that here we do not account for contamination
  (misclassified early- and late-type galaxies).}
\label{mrh}
\end{figure*}

\begin{figure*}[t]
\epsscale{.55} 
\plotone{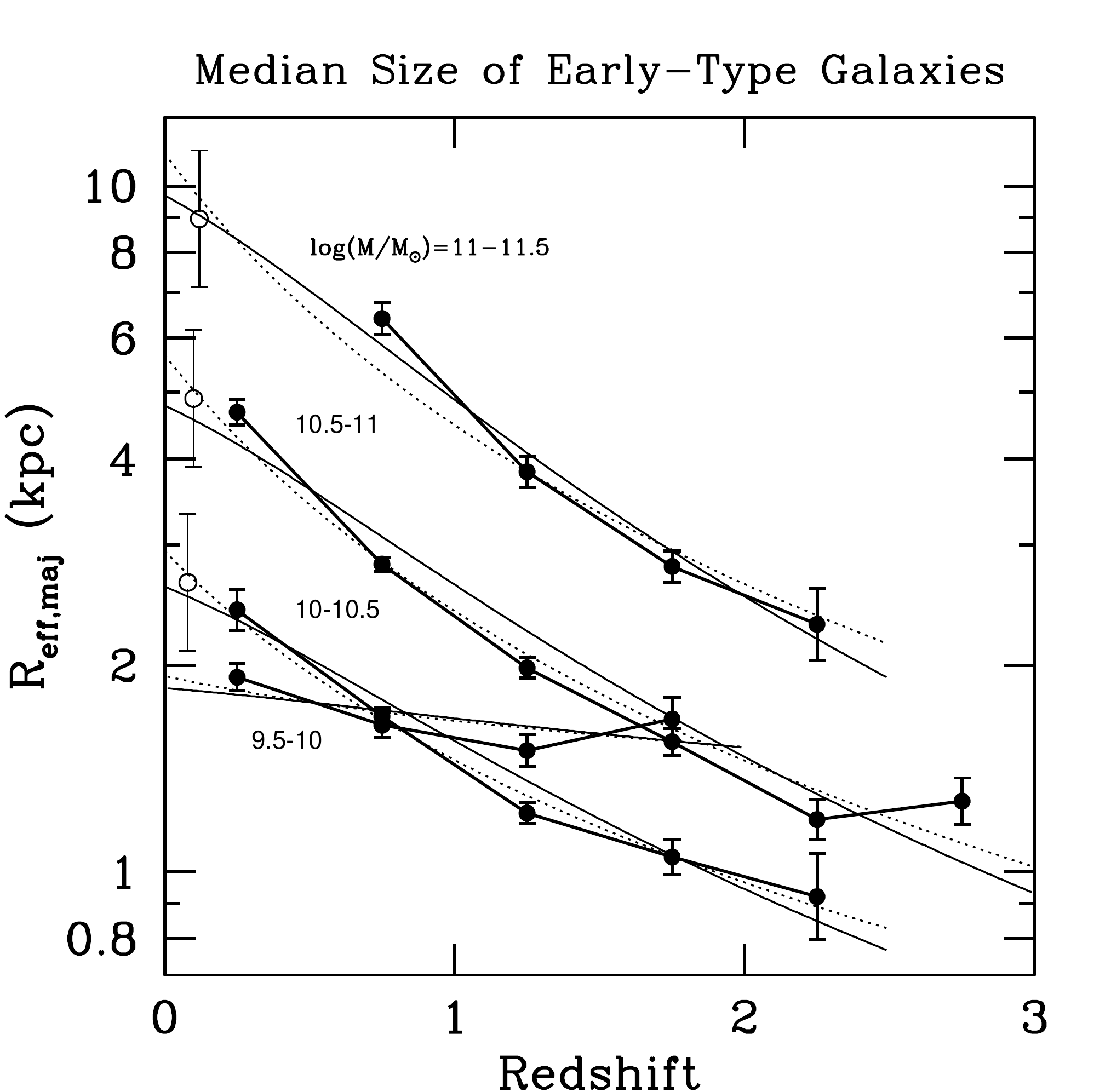}
\plotone{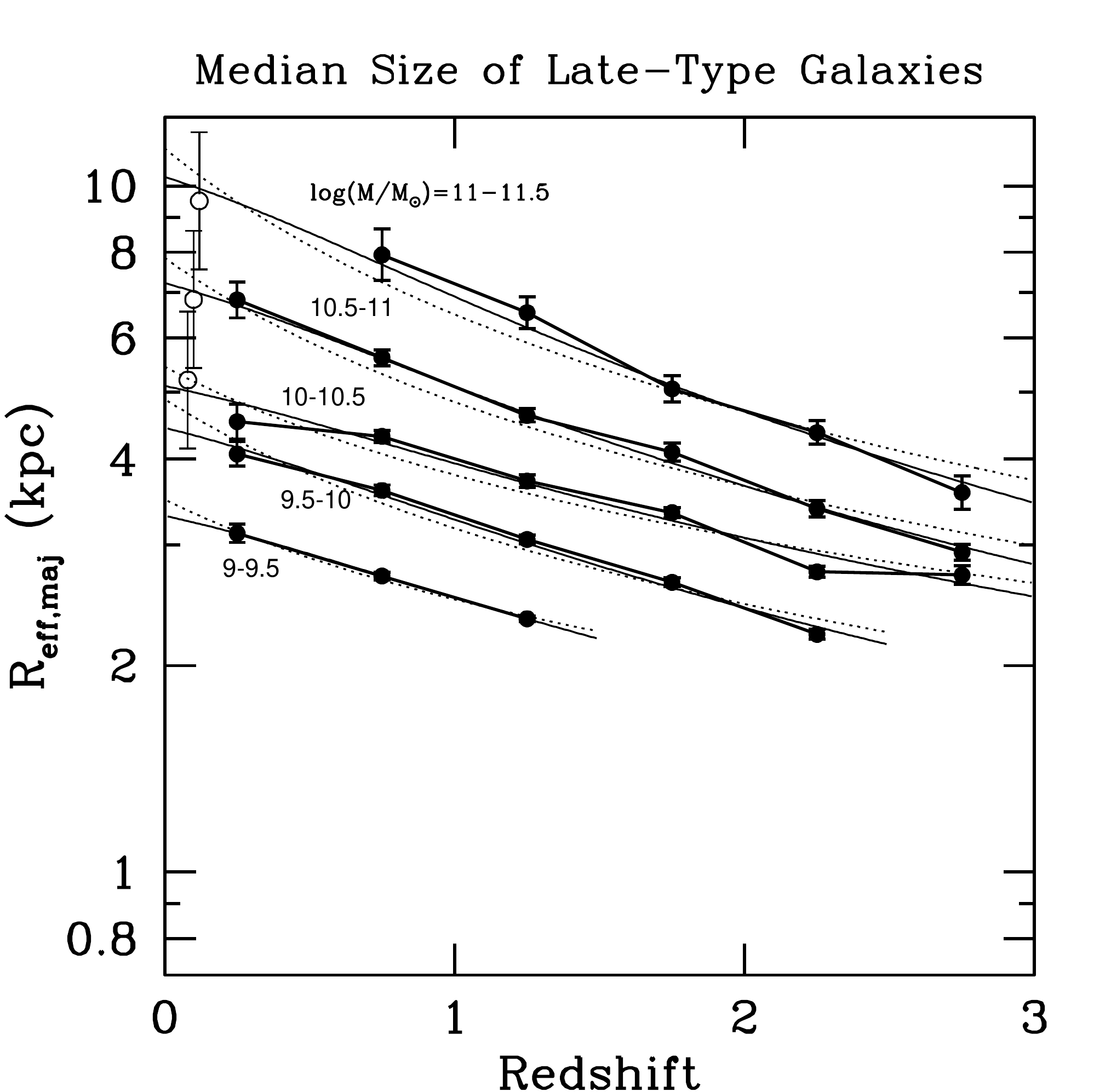}
\caption{Median size as a function of stellar mass and redshift for
  early-type galaxies (left) and late-type galaxies (right).  SDSS
  data points based on \citep{guo09} are shown as open points.  Fits
  to the median sizes of the form $\reff / \rm{kpc} =
  B_z(1+z)^{\beta_z}$ and $ B_H(H(z)/H_0)^{\beta_H}$ are shown by
  dotted and solid lines, respectively.  The evolution of the
  early-type galaxies is independent of mass at $M_*>2\times \msolb$:
  massive galaxies evolve fast and have a steep size-mass relation at
  all redshifts, while the relation flattens out at lower masses
  ($\lesssim 10^{10}~\msol$) and evolves less rapidly.  The evolution
  of the late-type galaxies is overall slower, and does not depend
  strongly on mass.  The low-mass early-type galaxies evolve at the
  roughly the same pace as the late-type galaxies. The median sizes
  and fitting results are given in Table \ref{tab:medsize}.}
\label{zr}
\end{figure*}

\begin{figure*}[t]
\epsscale{1.2} 
%\epsscale{0.5} 
\plotone{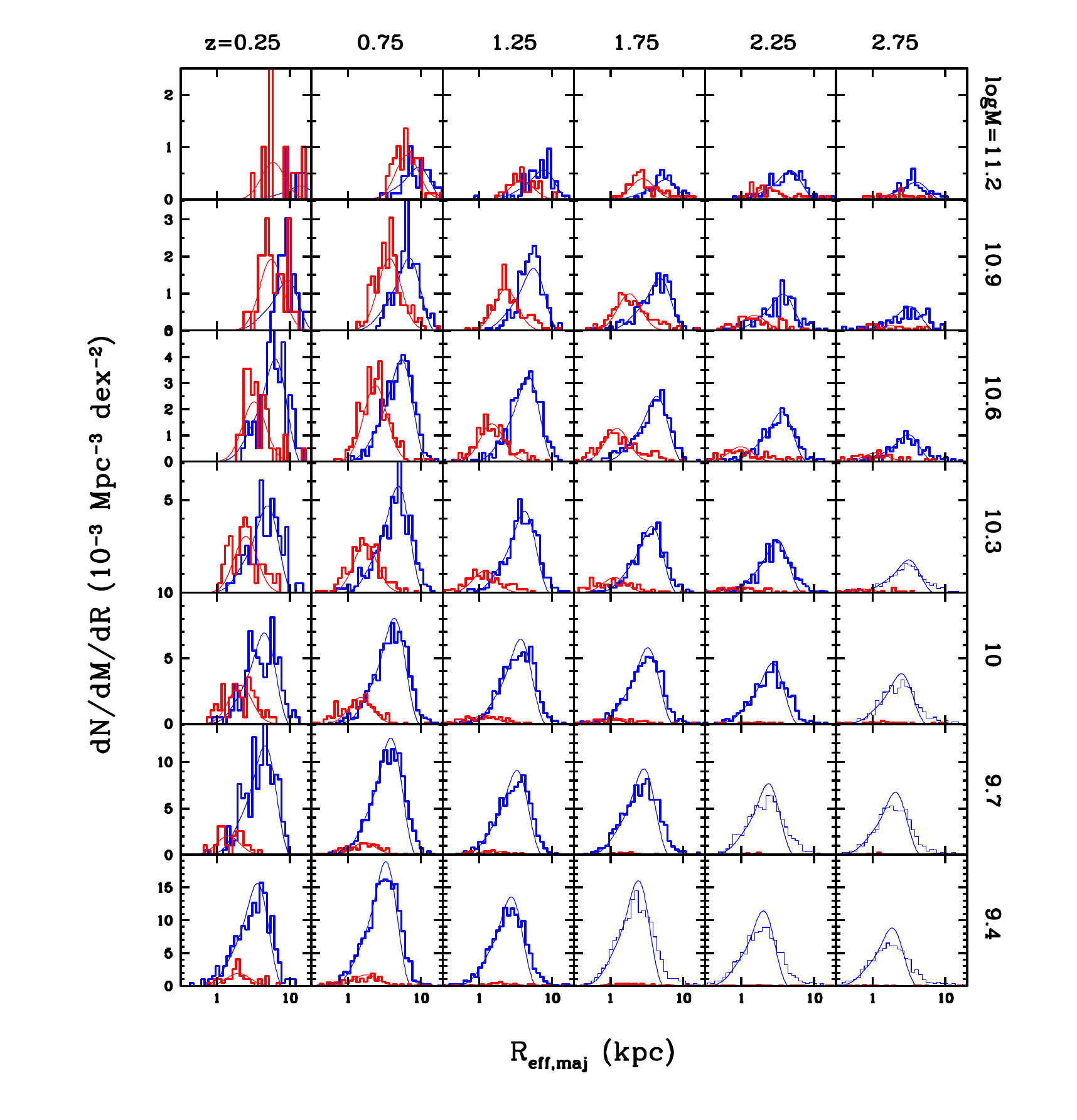}
\caption{Size distribution histograms for early- and late-type
  galaxies as a function of stellar mass (as labeled on the right-hand
  side) and redshift (as labeled at the top).  The number of galaxies
  is given in units of comoving volume to illustrate the growth of the
  population over time.  The early-type size distributions are fit
  with Gaussians with a fixed dispersion of 0.16 dex.  The late-type
  size distributions are fit with skewed Gaussians with a fixed
  dispersion of 0.16 dex and skewness $h_3=-0.15$.  The panels with
  thin lines show samples that are below our mass limit.}
\label{rhist}
\end{figure*}

\begin{figure}[t]
\epsscale{1.2} 
\plotone{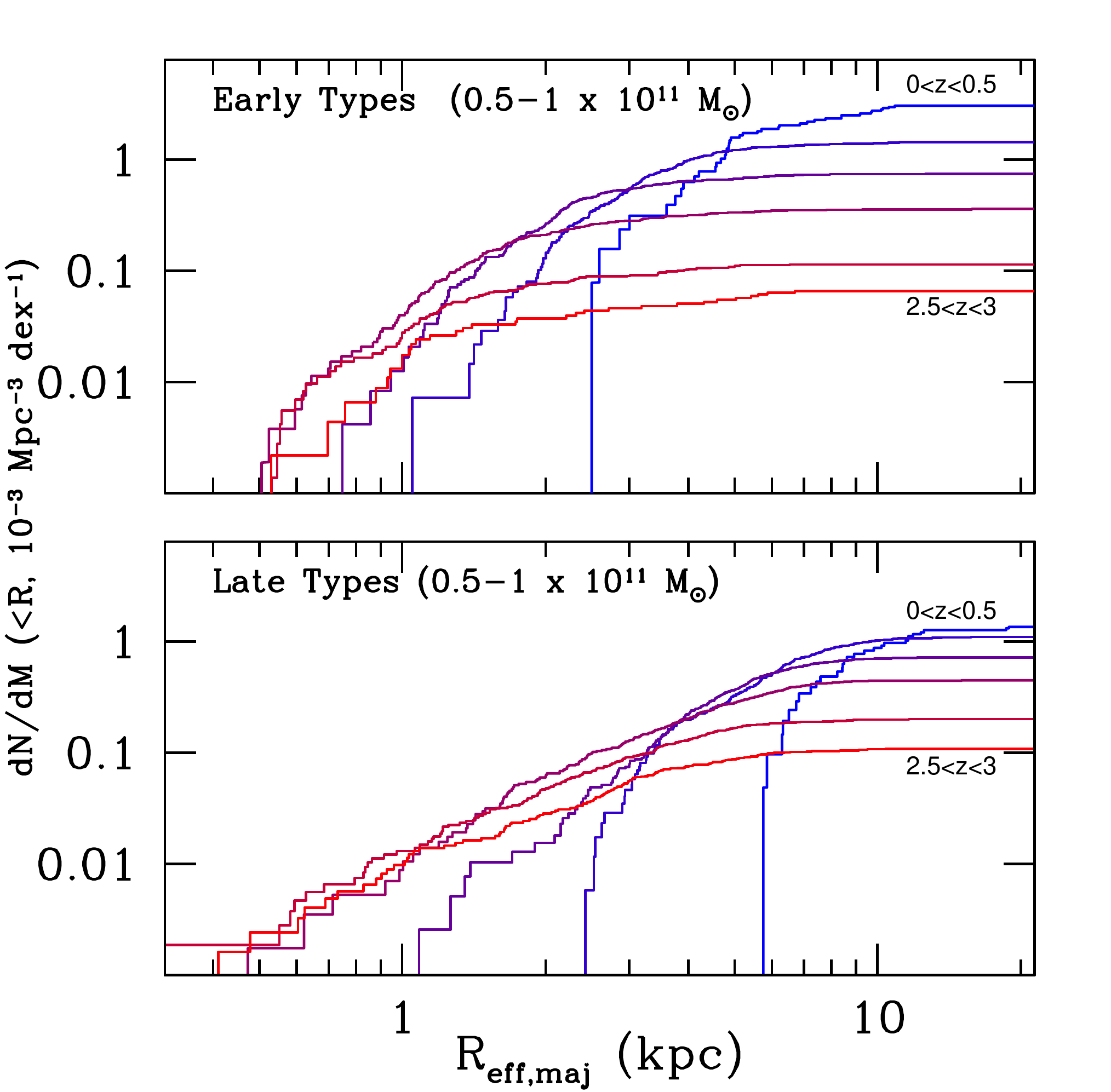}
\caption{Cumulative size distributions of $\sim L^*$ early-type
  galaxies (top) and $\sim L^*$ late-type galaxies (bottom) as a
  function of redshift.  While the number density of both early- and
  late-type galaxies increases over time, the number density of small
  galaxies declines, implying that the observed evolution in the mean
  size is not (solely) driven by the addition of larger galaxies.
  Individual galaxies must evolve in size.}
\label{rc}
\end{figure}

\begin{figure}[t]
\epsscale{1.2} 
\plotone{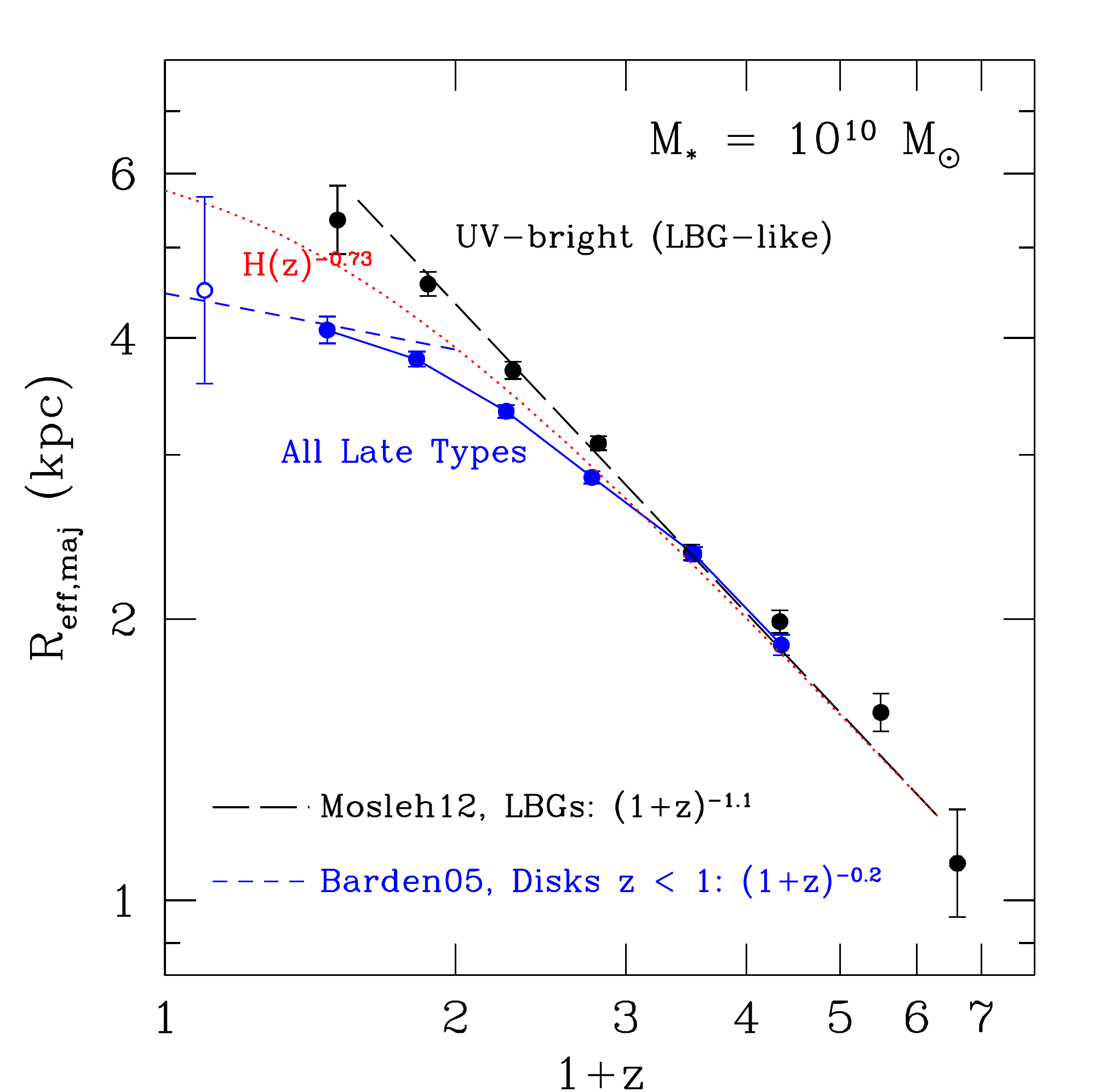}
\caption{Size evolution of galaxies in a narrow (0.3 dex) mass bin
  around $\msolb$.  The black points represent UV-bright galaxies
  (with $U-V<1$ in the rest-frame), selecting a sample akin to LBGs at
  high redshift.  Their size evolution is fast, consistent with the
  size evolution of UV-selected samples up to $z=7$ as recently
  determined by \citet{mosleh12} -- also see \citet{oesch10}. The blue
  points represent late-type galaxies as defined in this paper (see
  Figure \ref{uvj}), that is, all star-forming galaxies.  The size
  evolution of those is slower at low redshift, consistent with
  previous measurements at $z<1$ \citep[here,][]{barden05}.}
\label{lbg}
\end{figure}

\begin{figure}[t]
\epsscale{1.2} 
\plotone{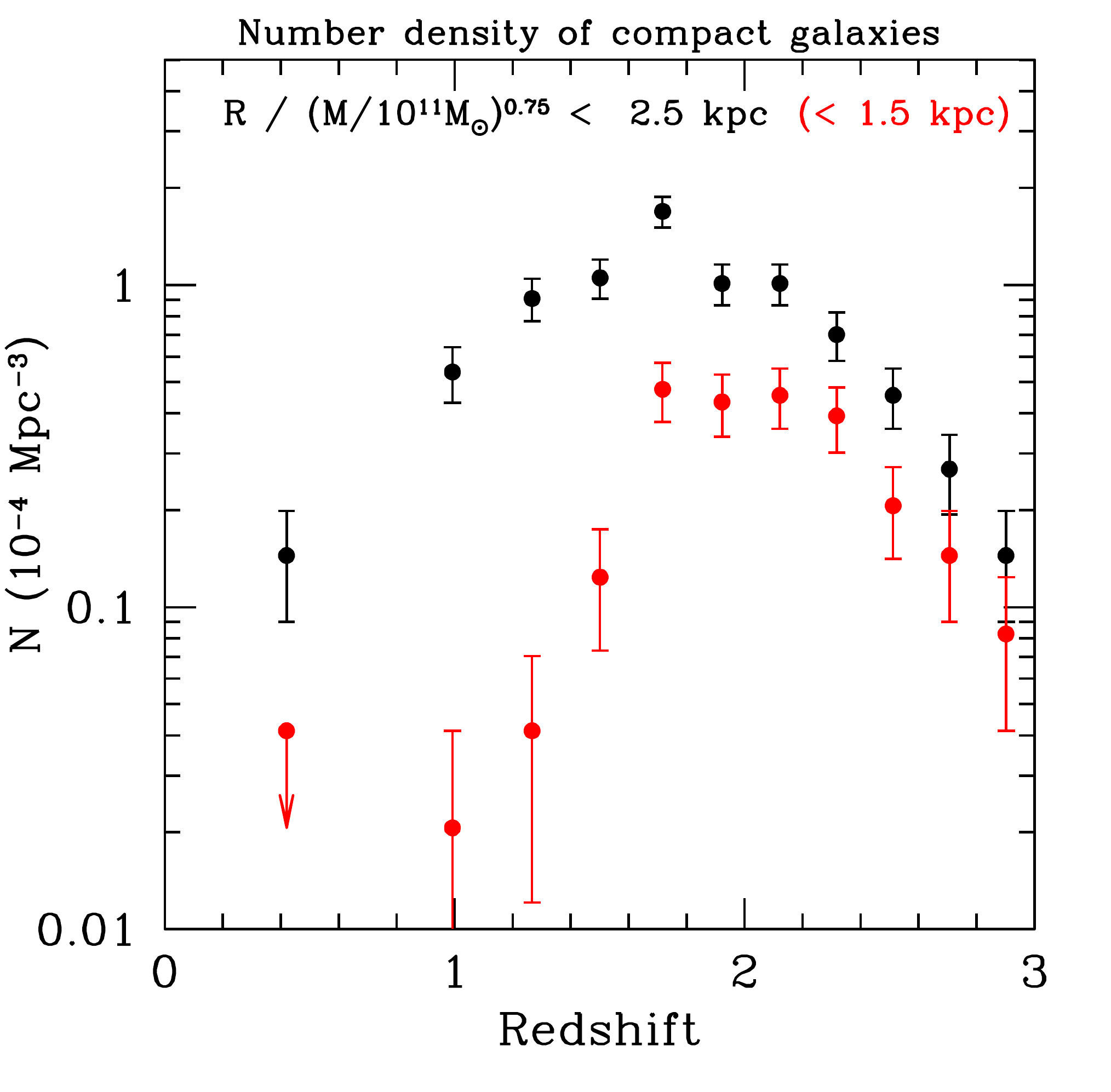}
\caption{Number density evolution of compact early-type galaxies.  In
  each redshift bin (with equal co-moving volume) we include
  early-type galaxies with mass $M_* > 5\times \msolb$ and size $\reff
  (\rm{kpc}) < (M_*/5\times \msolb)^{0.7}$.  That is, the slope of the
  size-mass relation is taken into account: for $M_* = 5\times \msolb$
  the size limit is $\reff = 1$~kpc and for $M_* = \msola$ the size
  limit is $\reff = 1.6$~kpc.  The number density first increases with
  cosmic time, reaching a plateau at $z\sim1.5-2$, after which it
  strongly decreases toward the present day.  The immediate
  implication is that individual galaxies must grow in size
  significantly, most likely through merging.}
\label{small}
\end{figure}

\section{Data}\label{sec:data}

The procedures for source detection, multiwavelength photometry,
redshift determinations, rest-frame color and stellar mass estimates
are described elsewhere.  Here, we briefly summarize these steps.

\subsection{Source Detection}\label{sec:det}
The 207,967 sources in all five CANDELS/3D-HST fields
\citep{koekemoer11, brammer12a} are detected in and extracted from
images that combine the available \emph{HST}/WFC3 IR channel data;
that is, stacked mosaics consisting of F125W, F140W, and F160W imaging
are constructed for this purpose.  We refer to the photometry data
release paper from the 3D-HST collaboration by \citet{skelton14} for
details.

%\subsection{Multi-Wavelength Photometry}\label{sec:phot}
\subsection{Photometric and Spectroscopic Redshift Determinations}\label{sec:z}

Multiwavelength photometry from \emph{HST}/WFC3 and \emph{HST}/ACS
imaging is produced by creating point spread function (PSF)-matched
images, using custom-made kernels and performing simple aperture
photometry.  Multiwavelength photometry from ground-based optical-NIR
and \emph{Spitzer}/IRAC imaging is produced using the approach
outlined by \citet{labbe06} and further developed by
\cite{gonzalez10}, which addresses blending by nearby sources and
takes the large differences in PSF into account through the use of
custom-made convolution kernels.  Like alternative methods such as
TFIT \citep{laidler07}, our approach uses high-resolution images as
priors to model sources in lower-resolution images.

Photometric redshifts are determined on the basis of the
multiwavelength photometry using the {\tt EAZY}~package
\citep{brammer08}.  \citet{skelton14} describe the procedure in full,
but it essentially follows \citet{whitaker11}: briefly, linear
combinations of a set of templates that span the range of observed
galaxy properties are used to fit the photometry, producing a
marginalized posterior probability distribution for the redshift, with
$z_{\rm{phot}}$ as its peak.

The photometric redshifts provide a baseline for 3D-HST WFC3 G141
grism spectroscopy to provide more precise redshift information.
\citet{brammer12a} describe the extraction of spectra and redshift
determination in detail.  The method has been updated to use
interlaced rather than drizzled \emph{HST}/WFC3 mosaics, which are
used to construct the photometric catalogs.  For all sources brighter
than $\rm{F}160\rm{W}_{\rm{AB}}=23$, the F140W image is traced along
the dispersed WFC3/G141 grism image, such that a spectrum is extracted
that accounts for the convolved spectral and spatial information of
the low wavelength resolution ($R\sim 130$) grism data.  For each
extracted object the spectroscopic information is combined with the
photometric redshift probability distribution, producing a new
best-fitting redshift (I.G.~Momcheva et al.~in preparation).  Finally,
spectroscopic redshifts from the literature are used when available.

The grism data significantly improve the redshift precision for
thousands of galaxies and provide indispensable evidence for the good
accuracy of the purely photometric redshift estimates.  The current
version of the 3D-HST redshift catalog contains grism redshift
information for all objects brighter than $\mh=23$ and for which such
data is available ($\sim$75\% of the CANDELS area).  For our
mass-limited sample -- defined below in Section \ref{sec:sample} --
this amounts to $\sim$10,000 galaxies.  This is $\sim$30\% of the
total sample and 50\% for the sample of massive ($M_*>\msolb$)
galaxies in the crucial redshift range $1<z<3$.  For these galaxies
the \citet{quadri10} pair test demonstrates a precision of $\Delta
z/(1+z)=0.003$, or 0.3\%.  For purely photometric redshifts this is
$1\%-2.5\%$, depending on the varying photometric data set available
for each of the five fields, suggesting a factor of $3-10$ improvement
in redshift precision from the grism data.  There is no systematic
offset between the two sets of redshifts.

\subsection{Rest-frame Colors and Stellar Mass Estimates}\label{sec:mass}

{\tt EAZY}~is used to compute the rest-frame $U-V$ and $V-J$ colors,
and the package {\tt FAST}~\citep{kriek09a} is used to estimate
stellar masses.  A large number of \citet{bruzual03} templates with
solar metallicity, a wide range in age ($4\times 10^7$~yr to
$12.5\times 10^9$~yr, but always younger than the universe),
exponentially declining star-formation histories (with time scales
$\tau=10^7 - 10^{10}$~yr), and dust extinction ($A_{\rm{V}}=0-4$) are
used and matched to the photometry.  The final stellar mass is
corrected for the difference between the total F160W flux from the
photometric catalog and the total F160W as measured with \gf (see
Section \ref{sec:size}).  F125W is used in case F160W is not
available.  This correction ensures that our size and mass estimates
are mutually consistent in that both are based on the same model for
the light distribution.

The uncertainties in the stellar mass estimates can be large and are
to some extent still unknown.  However, the possible error in the
mass-to-light ratio for low-mass blue galaxies with precisely known
redshifts due to uncertainties in the star formation history and
metallicity is not larger than a factor of a few.  Moreover, for
high-mass galaxies the relation between dynamical and stellar mass
estimates is the same to within a factor of $\sim 2$ over the whole
redshift range $0<z<2$, indicating that stellar mass estimates are
reasonably consistent at different redshifts
\citep[e.g.,][]{vanderwel06, vandesande13}.  The direct mass
measurement of a $z\sim 1.5$ strong gravitational lens is also in
agreement with the photometrically estimated stellar mass
\citep{vanderwel13}.  For these reasons we are confident that the
analysis presented in this paper, which looks at galaxies that span
several orders of magnitude in stellar mass, is not fundamentally
dependent on the current uncertainties in the mass estimates.

\subsection{Sample Selection}\label{sec:sample}

Following \citet{wuyts07} and \citet{williams09}, we utilize
rest-frame colors to distinguish between the two basic classes of
galaxies: star-forming and quiescent galaxies.  In this paper we refer
to the former as late-type galaxies and to the latter as early-type
galaxies.  In Figure \ref{uvj} we show the rest-frame $U-V$ and $V-J$
color distribution of galaxies with stellar mass in excess of $\msolb$
for a range of redshifts.  Even beyond $z=2$ the early- and late-type
galaxies are separated in this space, which allows us to effectively
assign each object to one of the two classes.

An indication of the high fidelity of this selection method is that
less than 20\% of the thus classified early-type galaxies with matches
in the FIREWORKS catalog from \citet{wuyts07} are detected at
$24\mu$m, while more than $\sim 80\%$ of the late-type galaxies are
detected -- these numbers are for the mass range $3\times \msolb <
\msol < \msola$ and the redshift range $1<z<2$.  We note that a
simpler selection by just $U-V$ color would compromise the separation
into types, as the subsample of galaxies that are red in $U-V$ consist
of dusty and quiescent objects.  In addition, 80\% of the
color-selected early-type (late-type) galaxies have S\'ersic indices
large (smaller) than $n=2.5$.

We adopt redshift- and color-dependent mass limits (Figure \ref{mmag})
that are set by the F160W magnitude limit down to which galaxy sizes
-- here described in Section \ref{sec:size} -- can be determined with high
fidelity.  \citet{vanderwel12} showed that galaxy sizes can be
precisely and accurately determined down to a magnitude limit of
$\mh=24.5$; at fainter magnitudes the random and systematic errors can
exceed 20\% for large galaxies with large S\'ersic indices.  Since
most $z>1$ galaxies in our sample have small sizes (0.3'' in the
median) and low S\'ersic indices (1.4 in the median) this magnitude
limit is conservative.

Out to $z=3$ our magnitude-limited sample is complete in stellar mass
down to $\sim \msolb$ (see Figure \ref{mmag}).  This limit is 1.9
magnitudes brighter than the $5\sigma$~detection limit, and
simulations of artificial objects inserted in the images show that
$\gtrsim 95\%$~of all objects are detected \citep{skelton14}.
Therefore, incompleteness will not be an issue for our sample, and
biases against large (or small) galaxies will not play a significant
role.

Our mass-selected sample, with a redshift-dependent mass limit as
described above (and a minimum of $M_* = \msolc$ and $\mh=25.5$ at all
redshifts), contains 32,722 galaxies.  43 of these are excluded
because of catastrophically failed surface brightness profile fits. We
then manually verified the spectral energy distributions (SEDs), size
measurements, mass and redshift estimates of all 7065 objects with
flagged \gf results \citep[$f=2$ from][]{vanderwel12}, small or large
sizes ($\reff<0.6$~kpc; $\reff>10$~kpc), large stellar masses
($M_*/\msol>10^{10.8}$), or large differences between photometric and
grism redshifts ($\Delta z > 0.15$), as well as all early-type
galaxies at redshifts $2 < z < 3$.  We removed 1721 (5.3\%) after this
verification because of errors in the size, redshift, or mass
measurements.  Hence, we have a final sample of 30,958 galaxies;
21,828 (5189) are at $z>1$ ($z>2$).

In Figure \ref{muv} we show the rest-frame $(U-V)$-$M_*$ distribution
and a clear bimodality is seen, equivalent to the bimodality seen in
Figure \ref{uvj}.  As is well-known, more massive galaxies are redder
and more likely to be quiescent, at least up to $z\sim 2.5$.  The most
massive galaxies $>\msola$ are essentially all red.  They are a mix of
quiescent and star-forming galaxies; the quiescent galaxies dominate
in number at $z<1$ and the star-forming galaxies dominate at $z>2$.

\subsection{Size Determinations}\label{sec:size}

Galaxy sizes are measured as described by \citet{vanderwel12}, from
mosaics created by the 3D-HST collaboration from public CANDELS F125W
and F160W and 3D-HST F140W raw imaging data.  With \gf we fit
single-component S\'ersic profiles to the two-dimensional light
profiles of all detected objects, making use of custom-made PSF
models; the package \gala allows for simultaneous fitting of as many
neighboring objects as needed, in order to avoid confusion.  As the
effective radius we use the semi-major axis of the ellipse that
contains half of the total flux of the best-fitting S\'ersic model.
For a full description of the procedure, we refer to
\citet{vanderwel12}; the size measurements published in their catalog
and the size measurements here\footnote{The results from our \gf
  profile fits used here, with IDs matched to the \citet{skelton14}
  catalogs, are publicly
  available for all 5 fields:\\
  \url{http://www.mpia-hd.mpg.de/homes/vdwel/3dhstcandels.html}.}  are
fully consistent.

Color gradients and their evolution affect the size measurements to a
degree that greatly exceeds the statistical uncertainties in our
sample averages and the size distributions as a function of galaxy
mass and redshift.  The dynamic range in wavelength for our sample is
rather small (from $1.25\mu$m to $1.6\mu$m), which does not allow us
to systematically probe the effect of color gradients over the large
range in redshift.  To extend the dynamic range, we analyze ACS/F814W
imaging in the COSMOS field from CANDELS parallel observations.  {\tt
  GALAPAGOS} is used to obtain galaxy sizes in precisely the same
manner for the F814W data as for the WFC3 data.

Using this extension in wavelength, we show in Figure \ref{mgrad} the
wavelength dependence of galaxy size for a sample of 777 late-type
galaxies.  The figure shows $\Delta \log \reff / \Delta \log \lambda$
as a function of galaxy mass and redshift.  For galaxies at $z<1$, the
difference between $\log \reff$ from the F814W and F125W size
measurements was taken as $\Delta \log \reff$; for $z>1$ galaxies the
difference between $\log \reff$ from the F125W and F160W measurements
was taken.  The pivot wavelengths of the respective filters are used
to compute $\Delta \log \lambda$.

The generally negative values of the quantity $\Delta \log \reff /
\Delta \log \lambda$ imply that late-type galaxies are typically
smaller at longer wavelengths and thus have negative color gradients
(redder centers).  The color gradients of $z>1$ galaxies have been
extensively studied before \citep[e.g.,][]{szomoru11, wuyts12}, and
here we merely mention the trends that are relevant for our conversion
of the measured $\reff$ to the $\reff$ at a rest-frame wavelength of
5000$\rm{\AA}$.  Color gradients are strongest for the most massive
galaxies at all redshifts and stronger at later cosmic times for
galaxies of all masses.  The evolution is remarkably smooth, which
allows us to parameterize the wavelength dependence of $\reff$ as a
simple function of redshift and galaxy stellar mass:

\begin{equation} \frac{\Delta \log \reff}{\Delta \log \lambda} = \\
-0.35  + 0.12 z -0.25 \log\Big(\frac{M_*}{\msolb}\Big).
\label{eq:grad}
\end{equation}
\citet{kelvin12} derive a very similar wavelength dependence of
$\reff$, but a direct comparison cannot be made as those authors do
not distinguish between galaxies with different masses.  Then, $\reff$
at a rest-frame wavelength of 5000$\rm{\AA}$ is estimated as

\begin{equation} \reff=R_{\rm{eff,F}} \\ 
\Big(\frac{1+z}{1+z_p}\Big)^{\frac{\Delta \log \reff}{\Delta \log \lambda}},
\end{equation}
where F denotes either F125W (for galaxies at $z<1.5$) or F160W (for
galaxies at $z>1.5$), and $z_p$ is the `pivot redshift' for these
respective filters (1.5 for F125W and 2.2 for F160W).  Positive color
gradients as computed with Equation (1) are set at zero.

A similarly detailed correction for early-type galaxies is currently
not feasible as that subsample is much smaller and has much redder
colors.  For the 122 early-type galaxies in the COSMOS field at
redshifts $0<z<2$ and with robust size measurements in all three
filters, we find an average size gradient of $\Delta \log \reff /
\Delta \log \lambda = -0.25$, with no discernible trends with mass and
redshift, which is in reasonable agreement with \citet{kelvin12} for
low-redshift early-type galaxies and \citet{guo11} for high-redshift
early-type galaxies.  Hence, we adopt this value for all early-type
galaxies when computing their rest-frame $5000\rm{\AA}$ sizes with
Equation 2.

This rather convoluted procedure has a small but significant effect on
the size measurements, with implications for the rate of size
evolution.  In the redshift range $1<z<2$, the size corrections are
always smaller than 0.05 dex because the observed wavelength is always
close to the desired rest-frame wavelength; at $z<1$ the correction
can increase up to 0.15 dex, but there the color gradient is well
constrained (see Figure \ref{mgrad}) for all galaxy masses.

\section{Evolution of the Size-Mass Distribution at $0 < z < 3$}\label{sec:distr}

The size-mass distributions for early- and late-type galaxies as a
function of redshift are shown in Figure \ref{mr}.  The first basic
observation is that a well-defined size-mass relation exists at all
redshifts $0<z<3$ for both populations.  With increasing redshift, the
two classes have increasingly different size-mass relations, mostly as
a result of strongly decreasing $\reff$ for early types.  In addition,
at all redshifts the slope of the relation is steeper for early types
than for late types.  At all stellar masses $\lesssim \msola$
early-type galaxies are, on average, smaller than late-type galaxies,
but at very high stellar masses ($>2\times\msola$) the two populations
have similar sizes at all redshifts, as far as the small samples allow
for such a comparison.

However, the relation for early-type galaxies flattens out below
$M_*=2\times 10^{10}~\msol$, at least up to $z=1.5$, beyond which our
sample is incomplete at these low masses.  This implies that the peak
in surface mass density at $\sim 4\times \msolb$ seen in the
present-day universe also existed at larger look-back times, at least
up to $z\sim 2$.

In Section \ref{sec:fitsize} we will first provide an analytical
description of the size-mass relation, which allows us to take
cross-contamination between the two types and outliers into account.
In the remainder of this section we will provide various direct
measurements such as median sizes and percentile distributions as a
function of mass and redshift and describe trends that are not
captured by our analytical description, such as deviations from a
single power law and skewness of the size distribution.

\subsection{Analytical Description}\label{sec:fitsize}
The basic characteristics of the galaxy size distribution are given by
the slope, intercept, and (intrinsic) scatter of size as a function of
mass.  We parameterize this following \citet{shen03} and assume a
log-normal distribution $N(\log r,\sigma_{\log r})$, where $\log r$ is
the mean and $\sigma_{\log r}$ is the dispersion.  Furthermore, $r$ is
taken to be a function of galaxy mass:

\begin{equation}r(m_*)/\rm{kpc}=A\cdot m_*^{\alpha},
\end{equation}
where $m_*\equiv M_*/7\times\msolb$.  As we will describe in Section
\ref{sec:skew}, it is reasonable to assume that $\sigma_{\log r}$ is
independent of mass.

The model distribution $N(\log r(m_*),\sigma_{\log r})$ prescribes the
probability distribution for observing $\reff$ for a galaxy with mass
$m_*$.  If the measured $\reff$ has a Gaussian, 1-$\sigma$ uncertainty
of $\delta \log \reff$, then the probability for this observation is
the inner product of two Gaussians:
\begin{equation}P = \langle N(\log \reff,\delta \log \reff),N(\log
  r(m_*),\sigma_{\log r})\rangle.\end{equation} Thus, we compute for
each galaxy the probabilities $P_{\rm{ET}}$ and $P_{\rm{LT}}$ for the
respective size-mass distribution models for the early-type and
late-type populations.  Incompleteness terms should formally be
included in these probabilities \citep[as described by,
e.g.,][]{huang13}, but because of our conservative sample selection
(see Section \ref{sec:sample}) we are not biased against faint, large
objects.

The uncertainty in size, $\delta \log \reff$, is computed as outlined
by \citet{vanderwel12}.  A random uncertainty of 0.15 dex in $m_*$ is
included in our analysis by treating it as an additional source of
uncertainty in $\reff$: for a size-mass relation with a given slope,
an offset in $m_*$ translates into an offset in $\reff$.  Hence, the
calculation of $P$ stays one-dimensional.  The fiducial slopes we use
to convert $\delta \log \reff$ into $\delta m_*$ are $\alpha=0.7$ for
early-type galaxies and $\alpha=0.2$ for late-type galaxies.

We also take into account the misclassification of early- and
late-type galaxies.  Despite the bimodal distribution in the
color-color diagram (Section \ref{sec:sample}; Figure \ref{uvj}),
there are galaxies in the region between the star-forming and
quiescent sequences, making their classification rather arbitrary and
causing cross-contamination of the two classes \citep[also
see][]{holden12}.  Motivated by this work, we take this
misclassification probability to be 10\%.  We will comment on the
effects of varying this parameter below, when we describe the fitting
results.

The misclassification probability precisely corresponds to the early-
and late-type contamination fractions in a sample in cases where the
two sub-samples have an equal number of galaxies.  The actual
contamination fraction scales with the early-, and late-type
fractions, which depend on galaxy mass and redshift.  The evolution of
the stellar mass function for the two types is described by
\citet{muzzin13}, which we use here to compute this ratio.  We also
allow for 1\% of outliers: these are objects that are not part of the
galaxy population, for example, catastrophic redshift estimates or
misclassified stars.  Finally, in order to avoid being dominated by
the large number of low-mass galaxies, we also assign a weight to each
galaxy thatq is inversely proportional to the number density. This
ensures that each mass range carries equal weight in the fit.  The
number density is taken from the \citet{muzzin13} mass functions.

Then, we compute the total likelihood for a set of six model
parameters (intercept $A$, slope $\alpha$, and intrinsic scatter
$\sigma_{\log\reff}$, each for both types of galaxies):
\begin{equation} \mathcal{L}_{\rm{ET}}=\sum \ln \Big[W \cdot \Big(
  (1-C) \cdot P_{\rm{ET}} + C \cdot P_{\rm{LT}} + 0.01 \Big)
  \Big]\end{equation}
for early-type galaxies, and
\begin{equation} \mathcal{L}_{\rm{LT}}=\sum \ln \Big[W \cdot \Big((1-
  C) \cdot P_{\rm{LT}} + C \cdot P_{\rm{ET}} + 0.01 \Big)
  \Big] \end{equation} for late-type galaxies, where $W$ is the weight
and $C$ is the contamination fraction, both of which are a function of
redshift and mass.  The best-fitting parameters are identified by
finding the model with the maximum total likelihood, $\mathcal{L} =
\mathcal{L}_{\rm{ET}}+\mathcal{L}_{\rm{LT}}$.

\begin{table*}[b]\scriptsize
\begin{center}
  \caption{ Results from the Parameterized Fits to the Size-Mass
    Distribution of the form $\reff / \rm{kpc} = A (M_*/5\cdot
    \msolb)^\alpha$, as Described in Section \ref{sec:fitsize} and
    Shown in Figures \ref{mr} and \ref{parevol}.
    $\sigma({\log{\reff}})$ is the scatter in $\reff$ in logarithmic
    units}
 \begin{tabular}{|c||c|c|c||c|c|c|}
\hline
& \multicolumn{3}{c||}{Early-type Galaxies} & \multicolumn{3}{c|}{Late-type Galaxies} \\
\cline{2-7}
 $z$ &   $\log(A)$ & $\alpha$ & $\sigma{\log({\reff}})$ &  $\log(A)$ & $\alpha$ & $\sigma{\log({\reff}})$ \\
\hline
0.25 & 0.60$\pm$0.02 & 0.75$\pm$0.06 & 0.10$\pm$0.02 & 0.86$\pm$0.02 & 0.25$\pm$0.02 & 0.16$\pm$0.01 \\
0.75 & 0.42$\pm$0.01 & 0.71$\pm$0.03 & 0.11$\pm$0.01 & 0.78$\pm$0.01 & 0.22$\pm$0.01 & 0.16$\pm$0.01 \\
1.25 & 0.22$\pm$0.01 & 0.76$\pm$0.04 & 0.12$\pm$0.01 & 0.70$\pm$0.01 & 0.22$\pm$0.01 & 0.17$\pm$0.01 \\
1.75 & 0.09$\pm$0.01 & 0.76$\pm$0.04 & 0.14$\pm$0.01 & 0.65$\pm$0.01 & 0.23$\pm$0.01 & 0.18$\pm$0.01 \\
2.25 &-0.05$\pm$0.02 & 0.76$\pm$0.04 & 0.14$\pm$0.02 & 0.55$\pm$0.01 & 0.22$\pm$0.01 & 0.19$\pm$0.01 \\
2.75 &-0.06$\pm$0.03 & 0.79$\pm$0.07 & 0.14$\pm$0.03 & 0.51$\pm$0.01 & 0.18$\pm$0.02 & 0.19$\pm$0.01 \\
\hline
\end{tabular}
\label{tab:fitsize}
\end{center}
\end{table*}

For the late types, we fit all galaxies with $M_*>3\times 10^9~\msol$;
this limit provides a good dynamic range of two orders of magnitude in
mass and exceeds the mass limit of our sample up to $z=2.5$ (Figure
\ref{mmag}).  For the early types, we fit all galaxies with
$M_*>2\times10^{10}~\msol$, so that we avoid the clearly flatter part
of the size-mass distribution at lower masses (see Section
\ref{sec:medsize}).  This cutoff exceeds the mass limit of our sample
up to $z=3$.

The black lines in Figure \ref{mr} indicate the fitting results, and
the evolution of the individual model parameters (intercept, slope,
and scatter) are shown in Figure \ref{parevol}.  The fitting results
are also given in Table \ref{tab:fitsize}.  The intercept of the
best-fitting size mass model distributions evolves significantly with
redshift and particularly rapidly for the early types.

Usually, the evolution of the intercept is parameterized as a function
of $(1+z)$.  While this is intuitively appealing because of our
familiarity with the cosmological scale factor, this is perhaps not
the physically most meaningful approach.  Galaxy sizes, in particular
disk scale lengths, are more directly related to the properties of
their dark matter halos than to the cosmological scale factor.  Halo
properties such as virial mass and radius follow the evolving
expansion rate -- the Hubble parameter $H(z)$ -- instead of the
cosmological scale factor.  For a matter-dominated universe, $H(z)$
and $(1+z)$ evolve at a similar pace, but as a result of the increased
importance at late times of $\Lambda$ for the dynamical evolution of
the universe, $H(z)$ evolves much slower in proportion to $(1+z)$ at
late times than at early times. For example, at $z\sim 0$ we have
$H(z)\propto (1+z)^{0.4}$, while at $z\sim 2$ this is $H(z)\propto
(1+z)^{1.4}$.

For this reason it is reasonable to parameterize size evolution as a
function of $H(z)$ in addition to $(1+z)$.  The solid lines in the
left-hand panel of Figure \ref{parevol} represent the evolution as a
function of $H(z)$, while the dashed lines represent the evolution as
a function of $(1+z)$.  These results are also given in Table
\ref{tab:fitsize}.  The $H(z)^{\beta_H}$ parameterization is
marginally preferred by the data over the $(1+z)^{\beta_z}$
parameterization, as is more clearly illustrated in Figure
\ref{zr_res}, where we show the residuals.  In addition to the
statistical limitations, we note that these residuals are of the same
magnitude as the systematic uncertainties in the size measurements and
color gradient corrections (Section \ref{sec:size}).  A more thorough
comparison with size evolution of larger samples at $z<1$ with size
measurements at visual wavelengths would improve these constraints.

\citet{newman12} first demonstrated the lack of strong evolution in
the slope of the size-mass relation for massive ($>2\times \msolb$)
early-type galaxies.  Here we confirm that result (middle panel,
Figure \ref{parevol}) and find a slope of $\reff\propto M^{0.75}$ at
all redshifts.  This slope is somewhat steeper than measured by
\citet{shen03} for present-day early-type galaxies.  Differences in
sample selection (star-formation activity versue concentration) and
methods ($\reff$ from S\'ersic profile fits versus Petrosian
half-light radii) may explain this difference. For the first time we
extend the analysis to late-type galaxies: the slope is much flatter
than the slope for early types ($\reff\propto M^{0.22}$), with little
or no change with redshift.  This slope is intermediate to the slope
found by \citet{shen03} for low- and high-mass galaxies.  Our sample
contains too few high-mass late-type galaxies to perform a robust
double-component power law fit, as done by \citet{shen03}, but in
Section \ref{sec:medsize} we will show evidence for the steepening of
the relation for massive late-type galaxies out to $z=1$.

Finally, we present the first measurement of the intrinsic scatter in
size beyond the local universe (right-hand panel of Figure
\ref{parevol}).  We find no strong evolution for either late types or
early types, and we find that the scatter for the early-type
population is always somewhat smaller (0.1 to 0.15 dex) than for the
late-type population (0.16 to 0.19 dex).  These numbers agree well
with the intrinsic scatter measured by \citet{shen03} for present-day
galaxies: 0.13 dex for early-type galaxies, and 0.20 dex for late-type
galaxies.  We note that the effects of measurement uncertainties were
not included by \citet{shen03}.

For comparison, we show the observed scatter at each redshift,
calculated as the standard deviation in $\reff$ after subtracting the
best-fitting size-mass relation.  The values for early-type galaxies
are in the range of 0.2-0.3 dex, in good agreement with the values
found by \citep{newman12} over the same redshift range.  In
particular, the strongly increased observed scatter in size for the
early-type galaxies at $z>2$ is largely attributed to significant
contamination by misclassified late-type galaxies.  We have assumed a
misclassification probability of $10\%$ (resulting in an assumed
misclassified fraction of $C=0.10$ in the case of an equal number of
early- and late-type galaxies -- see above), but although this value
is empirically motivated, it is not known with great precision.  If we
decrease (increase) the misclassification probability to 5\% (20\%),
then the recovered intrinsic scatter for the $z=2.0-2.5$ early-type
galaxy sample, for example, increases (decreases) to 0.18 (0.11).

At this point we should also comment on the effect of changing the
value for the assumed random uncertainty in stellar mass (here 0.15
dex).  Decreasing its value has no measurable effect, while increasing
it to 0.30 dex decreases the recovered value for the intrinsic scatter
further, to 0.05 dex for the $z=2.0-2.5$ early-type galaxy sample.  In
this sense, the derived values for the intrinsic scatter are upper
limits.

While our particular choices in modeling the uncertainties affect the
results with (marginal) significance, they do not affect our general
conclusions that the intrinsic size scatter (1) is $\lesssim$0.20 dex
for both types of galaxies and (2) does not strongly evolve with
redshift.  However, the conclusion that the scatter for early-type
galaxies is smaller than for late-type galaxies at all redshifts -- as
is seen for present-day galaxies -- should at this stage be regarded
as tentative.

Finally, we note that changes in the misclassification probability or
uncertainty in stellar mass do not significantly affect the recovered
values of the other model parameters (zero point and slope).

\subsection{Evolution of Median Sizes}\label{sec:medsize}

In this section we offer a complementary description of the evolution
of the size-mass relation.  In Figure \ref{mrh} we show the median
sizes as a function of mass and redshift, along with the 68th
percentile width of the size distribution.  The values are listed in
Table \ref{tab:medsize}.  Up to $z\sim1.5$ the relation for late types
steepens and tightens at the high-mass end.  \citet{shen03} modeled
the steepening by assuming a two-component power law, but we sample an
insufficiently large volume and sample size at $>\msola$ to include
this in our analytical description presented above (Section
\ref{sec:fitsize}).  The flattening of the size-mass relation for
low-mass early types is also clearly seen.  Inspection of the spectral
energy distributions of individual galaxies confirms that these are
truly quiescent galaxies, with strong 4000$\rm{\AA}$ breaks.  As we
showed in Section \ref{sec:fitsize}, the large apparent increase in
the scatter for high-redshift early types can be partially attributed
to contaminants and outliers.

We provide complementary sets of median size and scatter measurements
in the Appendix. These include the commonly used circularized radii:
$R_{\rm{eff,circ}} = \reff \sqrt{b/a}$, where $b/a$ is the projected
axis ratio.  In addition, we provide the measurements for the combined
late$+$early-type galaxy sample and the measurements in bins of
rest-frame $V$-band luminosity.

Figure \ref{zr} shows the median size evolution for galaxies in
different mass bins.  We parameterize this evolution both as a
function of $H(z)$ and of $(1+z)$ -- see Section \ref{sec:fitsize}.
The results are shown as solid and dotted lines, respectively, in
Figure \ref{zr} and are also given in Table \ref{tab:medsize}.

Ideally, an immediate comparison with the size-mass distribution of
nearby galaxies provides a strong constraint on the evolution.
However, such comparisons are fraught with systematic uncertainties.
The aim here is merely to show that our observations from CANDELS and
3D-HST are consistent with the size-mass relation for nearby galaxies
as measured from the SDSS \citep{shen03}, who provided the standard
reference for this purpose.

In order to account for possible systematic differences we compare the
size measurements from \citet{shen03} with those from \citet{guo09} on
an object-by-object basis.  The reason for using the \citet{guo09}
measurements as a baseline is that they are based on the same
technique -- \gala from \citet{barden12} -- as used in this paper.  We
shift the analytical descriptions of the size-mass relations from
\citet{shen03} according to the measured, systematic offset between
\citet{shen03} and \citet{guo09}.  This amounts to a 0.1 dex shift to
larger $\reff$ than \citet{shen03}.\footnote{Note that we use major
  axis $\reff$ in this paper, as opposed to \citet{shen03}, who use
  circularized radii} In order to be conservative, we also adopt 0.1
dex as the systematic uncertainty.  We show the inferred sizes for
local galaxies in Figure \ref{zr}.  The median size evolution traced
out by the 3D-HST/CANDELS data predicts $z\sim 0$ sizes that are
consistent with our adjusted \citet{shen03} median sizes.

The picture provided by median size distributions is consistent with
our analytical description (Section \ref{sec:fitsize}), with fast evolution
for the (massive) early types and moderate evolution for the late
types.  In addition, we see that the flattening of the relation for
low-mass early types coincides with slower evolution.  Interestingly,
low-mass early-type galaxies evolve at a similar same rate as late
types of the same mass.  For the late types we see a mild dependence
on mass: the more massive late types evolve slightly faster than the
less massive late types.  For further discussion, see
Section \ref{sec:discussion}.

\begin{table*}
\begin{scriptsize}
\begin{center}
  \caption{Logarithmic Size Distribution (16\% -- 84\% range) as a
    Function of Galaxy Mass and Redshift. The masses in the header and
    the redshifts in the first column are the centers of 0.5-wide
    bins. Redshift dependence in the form $\reff / \rm{kpc} =
    B_z(1+z)^{\beta_z}$ and $\reff / \rm{kpc} =
    B_H(H(z)/H_0)^{\beta_H}$ are also given. }
 \begin{tabular}{|cc||c|c|c|c|c||c|c|c|c|c|}
\hline
& & \multicolumn{5}{c||}{Early-Type Galaxies} & \multicolumn{5}{c|}{Late-Type Galaxies}\\
\cline{3-12}
%$M_{*,\odot}=10^{9.25}$
 $z$ &  &  {\tiny $M_*=10^{9.25}$}  & 9.75 & 10.25 & 10.75 & 11.25 & 9.25 & 9.75 & 10.25 & 10.75 & 11.25 \\
\hline
     & 16\% & 0.03$\pm$0.06 & 0.04$\pm$0.03 & 0.13$\pm$0.03 & 0.42$\pm$0.02 & 0.65$\pm$0.05 & 0.24$\pm$0.01 & 0.36$\pm$0.02 & 0.42$\pm$0.02 & 0.61$\pm$0.03 & \nodata \\ 
0.25 & 50\% & 0.27$\pm$0.02 & 0.28$\pm$0.02 & 0.38$\pm$0.03 & 0.67$\pm$0.02 & 0.76$\pm$0.09 & 0.49$\pm$0.01 & 0.61$\pm$0.02 & 0.66$\pm$0.03 & 0.83$\pm$0.03 & \nodata \\ 
     & 84\% & 0.46$\pm$0.03 & 0.46$\pm$0.04 & 0.58$\pm$0.04 & 0.92$\pm$0.05 & 1.08$\pm$0.08 & 0.70$\pm$0.01 & 0.80$\pm$0.01 & 0.85$\pm$0.02 & 1.01$\pm$0.04 & \nodata \\ 
\hline
     & 16\% &-0.02$\pm$0.03 &-0.14$\pm$0.02  & 0.02$\pm$0.02 & 0.26$\pm$0.01 & 0.62$\pm$0.02 & 0.18$\pm$0.01 & 0.32$\pm$0.01 & 0.39$\pm$0.01 & 0.51$\pm$0.02 & 0.77$\pm$0.02 \\ 
0.75 & 50\% & 0.23$\pm$0.01 & 0.21$\pm$0.02  & 0.23$\pm$0.01 & 0.45$\pm$0.01 & 0.81$\pm$0.02 & 0.43$\pm$0.01 & 0.56$\pm$0.01 & 0.64$\pm$0.01 & 0.75$\pm$0.01 & 0.90$\pm$0.04 \\
     & 84\% & 0.43$\pm$0.02 & 0.44$\pm$0.02  & 0.42$\pm$0.02 & 0.64$\pm$0.02 & 0.97$\pm$0.02 & 0.65$\pm$0.01 & 0.76$\pm$0.01 & 0.83$\pm$0.01 & 0.90$\pm$0.01 & 1.12$\pm$0.03 \\
\hline
     & 16\% & \nodata & -0.15$\pm$0.03 & -0.15$\pm$0.02 & 0.07$\pm$0.01 & 0.41$\pm$0.02 & 0.11$\pm$0.01 & 0.23$\pm$0.01 & 0.33$\pm$0.01 & 0.47$\pm$0.01 & 0.62$\pm$0.03 \\
1.25 & 50\% & \nodata & 0.18$\pm$0.03 & 0.09$\pm$0.02 & 0.30$\pm$0.01 & 0.58$\pm$0.03 & 0.37$\pm$0.01 & 0.48$\pm$0.01 & 0.57$\pm$0.01 & 0.67$\pm$0.01 & 0.82$\pm$0.03 \\
     & 84\% & \nodata & 0.42$\pm$0.04 & 0.36$\pm$0.03 & 0.54$\pm$0.03 & 0.81$\pm$0.03 & 0.60$\pm$0.01 & 0.69$\pm$0.00 & 0.77$\pm$0.01 & 0.83$\pm$0.01 & 0.96$\pm$0.02 \\
\hline
     & 16\% & \nodata & -0.02$\pm$0.06 & -0.27$\pm$0.02 & -0.04$\pm$0.02 & 0.28$\pm$0.02 & 0.07$\pm$0.01 & 0.16$\pm$0.01 & 0.28$\pm$0.01 & 0.35$\pm$0.02 & 0.53$\pm$0.04 \\
1.75 & 50\% & \nodata &  0.22$\pm$0.03 & 0.02$\pm$0.03 & 0.19$\pm$0.02 & 0.45$\pm$0.02 & 0.33$\pm$0.01 & 0.42$\pm$0.01 & 0.52$\pm$0.01 & 0.61$\pm$0.01 & 0.70$\pm$0.02 \\
     & 84\% & \nodata &  0.48$\pm$0.06 & 0.35$\pm$0.03 & 0.50$\pm$0.03 & 0.74$\pm$0.04 & 0.57$\pm$0.01 & 0.65$\pm$0.01 & 0.72$\pm$0.01 & 0.80$\pm$0.01 & 0.87$\pm$0.02 \\
\hline
     & 16\% & \nodata & \nodata & -0.37$\pm$0.08 & -0.20$\pm$0.02 & 0.16$\pm$0.03 & \nodata & 0.10$\pm$0.01 & 0.17$\pm$0.02 & 0.26$\pm$0.03 & 0.40$\pm$0.02 \\
2.25 & 50\% & \nodata & \nodata & -0.04$\pm$0.07 & 0.08$\pm$0.03  & 0.36$\pm$0.05 & \nodata & 0.35$\pm$0.01 & 0.44$\pm$0.01 & 0.53$\pm$0.01 & 0.64$\pm$0.02 \\
     & 84\% & \nodata & \nodata & 0.36$\pm$0.03  & 0.54$\pm$0.04  & 0.55$\pm$0.07 & \nodata & 0.57$\pm$0.01 & 0.64$\pm$0.01 & 0.70$\pm$0.01 & 0.84$\pm$0.02 \\
\hline
     & 16\% & \nodata & \nodata & \nodata & -0.22$\pm$0.05 & 0.07$\pm$0.07 & \nodata & \nodata & 0.16$\pm$0.02 & 0.19$\pm$0.06 & 0.33$\pm$0.04 \\
2.75 & 50\% & \nodata & \nodata & \nodata & 0.10$\pm$0.03  & 0.39$\pm$0.08 & \nodata & \nodata & 0.43$\pm$0.01 & 0.47$\pm$0.01 & 0.55$\pm$0.02 \\
     & 84\% & \nodata & \nodata & \nodata & 0.50$\pm$0.10  & 0.68$\pm$0.10 & \nodata & \nodata & 0.65$\pm$0.02 & 0.71$\pm$0.02 & 0.76$\pm$0.04 \\
\hline
\hline
 & $\log(B_z)$     & \nodata &  0.29$\pm$0.01 & 0.47$\pm$0.02  & 0.75$\pm$0.04  &  1.05$\pm$0.09 & 0.54$\pm$0.01  & 0.69$\pm$0.01  & 0.74$\pm$0.03  & 0.90$\pm$0.05  & 1.05$\pm$0.08 \\
 & $\beta_z$ & \nodata & -0.22$\pm$0.02 & -1.01$\pm$0.06 & -1.24$\pm$0.08 & -1.32$\pm$0.21 & -0.48$\pm$0.03 & -0.63$\pm$0.02 & -0.52$\pm$0.08 & -0.72$\pm$0.09 & -0.80$\pm$0.18 \\
\hline
 & $\log(B_H)$     & \nodata &  0.27$\pm$0.01 & 0.42$\pm$0.02  & 0.68$\pm$0.03  &  0.97$\pm$0.06 & 0.52$\pm$0.01  & 0.65$\pm$0.01  & 0.71$\pm$0.03  & 0.86$\pm$0.04  & 1.01$\pm$0.06 \\
 & $\beta_H$ & \nodata & -0.19$\pm$0.02 & -0.97$\pm$0.05 & -1.13$\pm$0.06 & -1.29$\pm$0.16 & -0.52$\pm$0.02 & -0.58$\pm$0.02 & -0.49$\pm$0.07 & -0.65$\pm$0.09 & -0.76$\pm$0.13 \\
\hline
\end{tabular}
\label{tab:medsize}
\end{center}
\end{scriptsize}
\end{table*}

\subsection{Skewness in the $\reff$ Distribution of Late-type
  Galaxies}\label{sec:skew}

The 16- and 84-percentile range for late-type galaxies is not
precisely centered on the median size (Figure \ref{mrh} and Table
\ref{tab:medsize}), implying a skewness in the size distribution.  To
examine this further, we show size distribution histograms for a set
of mass and redshift bins in Figure \ref{rhist}.  The asymmetric size
distribution for late-type galaxies is due to a tail of small
galaxies.  The small sizes are not due to pointlike contributions from
an Active Galactic Nucleus (AGN): the $1-10\mu$m photometry of these
objects does not show the powerlaw SEDs that are characteristic for
unobscured AGN.

As a result of this skewness, there is substantial overlap between the
size distributions of early types and late types and no clear
bimodality despite large differences in the average sizes.  Figure
\ref{rc} shows more clearly than Figure \ref{rhist} that the size
distributions of the two types overlap at all redshifts, up to $z\sim
3$.

Figure \ref{rhist} also shows tails of large early-type galaxies.
However, this can likely be attributed to misclassification: their
number is always an order magnitude smaller than the number of
late-type galaxies with similar sizes, consistent with the assumed
misclassification probability in our analysis in
Section~\ref{sec:fitsize}.  We note that the tail of small late-type
galaxies is not consistent with the expected number of
misclassificatied objects: its prominence appears to be unrelated to
the early-type population.

As an illustration that the size distribution for both early- and
late-type galaxies evolve smoothly and regularly, we fit
Hermite-Gaussian functions to the histograms shown in Figure
\ref{rhist}.  This provides a reasonable description of all redshift
and mass bins.

For a discussion of the implications of these results in the context
of previous results, we refer to Sections \ref{sec:compearly} and
\ref{sec:early}.

\section{Comparison with Previous Results}\label{sec:comparison}

\subsection{Late-type Galaxies}\label{sec:complate}

The results for size evolution of late-type galaxies shown in Figures
\ref{parevol} and \ref{zr} are consistent with most other recent
measurements that focus on the $z=1-2$ redshift range
\citep[e.g.,][]{franx08, buitrago08}.  \citet{williams10} found
somewhat faster evolution for late types (and slower evolution for
early types), but a direct comparison with their Figure 4 reveals that
their size measurements and our measurements are in fact fully
consistent over the redshift range $0.5 < z < 1.5$.  The difference in
the quoted pace of evolution is likely the result of the increased
dynamic range in redshift probed here, in addition to the use of a
low-redshift comparison sample from the SDSS (see Section
\ref{sec:medsize}).

Thus, like previous studies focusing on the $z=1-2$ redshift regime,
we find that the pace of evolution for late-type galaxies is
intermediate to the slow evolution of disk galaxies at $z<1$
\citep{lilly98, ravindranath04, barden05} and the fast evolution of
UV-selected galaxies at $z>2$ \citep{giavalisco96, ferguson04,
  oesch10, mosleh12}.  Our data set allows us to bridge these regimes
and probe the origin of this difference.

In Figure \ref{lbg} we show the size evolution of galaxies with
stellar mass $M_*\sim \msolb$ from the present day up to $z\sim 6$.
Here we have relaxed our magnitude limit to $\mh=26$, which is still
within the completeness limit of the CANDELS imaging as can be seen in
Figure \ref{mmag}.  Size measurements of individual galaxies are no
longer reliable at $\mh=26$, but the sample average is still robust to
within 15\% \citep{vanderwel12}.  Using a color cut of $U-V < 1$ we
can probe a population akin to LBGs out to $z\sim 6$.  The median size
evolves quickly with redshift, $\reff\propto (1+z)^{-1.1}$, consistent
with recent measurements by \citet{oesch10} and \citet{mosleh12}.

Once we include all late-type galaxies, regardless of color, the
evolution matches that of the $U-V<1$ galaxies at $z\gtrsim 2$.  This
is simply because essentially all galaxies are blue: the color-blind
sample is not biased against galaxies with $U-V>1$ up to $z=3-3.5$.
At lower redshift, red star-forming galaxies appear, which are smaller
in size than their blue counterparts and slow down the average size
evolution.  At $z<1$, UV-bright galaxies are rare and the evolution is
dominated by redder galaxies, which evolve in size more slowly in
agreement with the results from, for example, \citet{barden05}.

As we argued in Section \ref{sec:medsize}, galaxy sizes may be better
parameterized as a function of $H(z)$ than as a function of $(1+z)$.
While the former naturally implies slower evolution at late times than
the latter (see the red dotted line in Figure \ref{lbg}), not all
trends are captured by using the $H(z)$ parameterization: (1) the
evolution of all late types slows down more rapidly than can be
explained by the difference between the two parameterizations, and (2)
the UV-bright sample shows little evidence for slowed evolution at
$z\lesssim 1$.

We conclude that the diverging pace of evolution seen at $z<1$ and
$z>3$ as reported in the literature is partly due to sample selection
effects and partly due to the different evolution of red and blue
late-type galaxies.

\subsection{Early-type Galaxies}\label{sec:compearly}

As we discussed in the Introduction, there is broad agreement that the
average sizes of $\sim L^*$ early-type galaxies, as measured at a
fixed stellar mass, were smaller at high redshift.  Moderate
deviations in the pace of evolution can be attributed to sample
selection, measurement and/or fitting techniques.  For example, as we
mentioned above, whereas the size measurements reported by
\citet{williams10} are consistent with our size measurements, the
reported pace of evolution is somewhat different: $(1+z)^{-1}$ from
\citet{williams10} and $(1+z)^{-1.3}$ from Section \ref{sec:medsize}
in this paper. This is the result of the difference in spanned
redshift range and the different use of present-day comparison
samples.  While these differences are large enough to be interesting,
there is a reasonable consensus that the average size for the
population of early-type galaxies evolves rapidly.  In particular, the
first \emph{HST}/NICMOS-based studues produced an impressive body of
evidence for rapid evolution \citep{zirm07, toft07, vandokkum08,
  buitrago08}, later confirmed by \emph{HST}/WFC3-based studies
\citep[e.g.,][]{newman12, cassata13, morishita14}.

What has so far remained contentious is what drives this evolution:
the size evolution of individual galaxies, the addition of larger
galaxies to the population, or a combination of both.  Figure \ref{rc}
(top panel) shows the cumulative size histograms of early-type
galaxies in the $L^*$ mass range, which reveal that the number density
of small early-type galaxies strongly evolves with redshift.  The
total number density of early-type galaxies increases from $z\sim 3$
to the present, but the number of small galaxies strongly decreases at
late cosmic times.  We show this explicitly in Figure \ref{small}: the
number density of high-mass galaxies with small sizes increases from
early times to $z\sim 2$, levels off, and then decreases strongly at
$z\lesssim 1.5$.  The definition of small is arbitrary, but the
general picture does not depend on the precise choice in mass and size
range.  This finding is in general agreement with previous claims
based on smaller samples and fewer fields by \citet{cassata11},
\citet{newman12}, \citet{szomoru12}, \citet{buitrago13}, and
\citet{cassata13}.

Several authors have argued that there are a substantial number of
small yet massive galaxies in the present-day universe
\citep{valentinuzzi10, poggianti13}.  The latter show that $3\%-5\%$
of present-day group and cluster early-type galaxies with mass
$M_*>3\times \msolb$ can be classified as ``compact.''  Following
their definition ($M_*/(2\pi R_{\rm{eff,circ}})>3\times
10^9~\msol~\rm{kpc}^{-2}$), we find that $\sim 40\%-50\%$ of $z\sim 1$
early-type galaxies qualify as compact.  Since the total number
density of such galaxies has evolved by no more than a factor two or
three we conclude that most of the $z\sim 1$ compact galaxies no
longer exist in that form in the present-day universe.  Several
``fossils'' in the form of $\reff\sim 1$ kpc, $M_*\sim \msola$
galaxies are found in the local volume \citep[see, e.g.,][for recent
examples]{vandenbosch12, dullo13}, but their number density is too low
to match the number density of their $z=2$ counterparts.

Recently, \citet{carollo13} claimed that the number density of small
early-type galaxies in the $L^*$ mass range has not strongly evolved
since $z\sim 1$.  We rule out that field-to-field variations explain
the discrepancy with our results.  All five fields show a decline in
the number density of compact galaxies (as defined in Figure
\ref{small}, with $\reff < 2.5$ at $\msola$) between $z=1.5$ and
$z=0.5$, by factors ranging from 3 to 10.  A decline of more than a
factor two between $z=1$ and $z=0.5$ is seen for four out of five
fields.

There are several factors, in the form of systematic effects in the
size and mass measurements, that may contribute to this tension.
Slight redshift-dependent shifts in the stellar mass estimates produce
changes in the size distribution as measured at a fixed mass.  Our
stellar mass estimates for luminous early-type galaxies have been
demonstrated to show small, if any, shifts ($\lesssim$0.1~dex)
compared with dynamical mass measurements over the redshift range
$0<z<2$ \citep{vandesande13,bezanson13,belli14}.  In addition, our
color-gradient correction would introduce a 14\% shift of $z=1$ sizes
relative to $z=0.2$ sizes in the Carollo et al.~sample.

Most importantly, the size measurements used here and by
\citet{carollo13} are obtained with fundamentally different
techniques: here we use parameterized profile fits, while Carollo et
al.~use growth curves.  The growth curve method does not take the PSF
into account at the time of measurement, but it relies on \textit{a
  posteriori} correction.  The magnitude of the correction depends on
the intrinsic structural properties and is inferred from simulated
size measurements.  For example, galaxies with measured sizes of
$\sim$0.\arcsec2~receive a negligible correction, whereas galaxies
with measured sizes of $\sim$0.\arcsec1~receive a factor of 2
correction downward (see Figure 2 of Carollo et al.).  With such
strongly size-dependent corrections it is difficult to reconstruct the
true size distribution, especially when those corrections are of a
similar magnitude as the sizes themselves.

In an explicit example in which we apply a systematic,
redshift-dependent shift in $\reff$ of order 0.1-0.2 dex per unit
redshift we infer non-evolution in the number density of compact
galaxies.  Given this sensitivity to small shifts in size, we argue
that our measurements, which do not require systematic size
corrections of more than a few percent (see Section \ref{sec:size} and
\citet{vanderwel12}), represent the size distribution with good
fidelity across the examined redshift range.

\section{Discussion}\label{sec:discussion}

\subsection{Evolution of Late-type Galaxies}\label{sec:late}

Remarkably, the observed pace of size evolution for late-type galaxies
is essentially the same as the evolution of the dark matter halo
radius at a fixed mass, $R\propto H(z)^{-2/3}$, but only when halo
mass and radius are defined with respect to the critical density.  In
a $\Lambda$CDM universe, if halo mass is parameterized with respect to
matter density or virial density (assuming top-hat collapse), then
$\Lambda$ causes strong departures from a power law at late cosmic
times.  The average evolution between $z=2$ and the present is
$R\propto H(z)^{-1.06}$ or $\propto H(z)^{-1.24}$, respectively
\citep{peebles80}.

The interpretation of such a comparison is not straightforward.
However, our novel measurement of the slope and scatter of the size
mass relation provides new constraints.  The intrinsic scatter in
galaxy size remains approximately the same at all redshifts ($\sim
0.16-0.19$~dex, see Figure \ref{parevol}) and is comparable to, but
perhaps somewhat smaller than, the scatter of 0.25 dex in the halo
spin parameter \citep[e.g.,][]{maccio08}.  This strongly suggests that
at all redshifts the sizes of late-type galaxies are set by their dark
matter halos, and it encourages us to examine the relation between
galaxy sizes and halo properties further.

The power law fits presented in Section \ref{sec:fitsize} imply that
there is very little or no evolution in the slope of the size-mass
relation; it remains flat, $\alpha \equiv d \log{\reff}/d
\log{M_*}=0.22\pm0.03$, at all redshifts $0<z<3$ (Figure
\ref{parevol}, middle panel).  As argued by \citet{shen03}, the flat
slope suggests that the ratio between galaxy mass and halo mass is not
a constant: if it were, the size-mass relation would be steeper
($\alpha=1/3$) than observed.  The underlying assumption is that
galaxy size is proportional to halo size \citep{kravtsov13}, which we
here take to be the case for late-type galaxies.  Following
\citet{shen03}, we use the observed slope ($\alpha\sim 1/5$) to
constrain the galaxy mass-halo mass relation and find $m_g\equiv
M_{\rm{gal}}/M_{\rm{halo}}\propto M_{\rm{halo}}^{\gamma\sim 2/3}$.
The observation that the slope of the size-mass relation does not
evolve with redshift provides a very stringent constraint on the
models: unless a combination of factors conspire to keep this slope
constant, the most straightforward explanation is that the slope of
the relation between galaxy and halo mass ($\gamma$) is similar across
the redshift range considered here.

Indeed, entirely independent estimates of the relationship between
galaxy and halo mass, based on clustering measurements and abundance
matching techniques, provide strong evidence that $m_g$ depends on
halo mass, similarly so at different redshifts
\citep[e.g.,][]{conroy09,moster10,behroozi10,wake11,moster13,behroozi13}.
In fact, the most recent studies found that $\gamma=2/3$ for halos in
the mass range $M_{\rm{halo}}\sim 10^{11-12}~\msol$, in agreement with
what we infer on the basis of the slope of the size-mass relation.  In
addition, \citet{moster13} and \citet{behroozi13} showed that $m_g$
peaks at a similar halo mass ($\sim 10^{12}~\msol$) at all redshifts
$z\lesssim 2$, at around a constant value of $m_g\sim 0.05$.  The
implication is that $m_g$ does not strongly evolve over the (rather
narrow) halo mass range $10^{11-12}~\msol$ \citep{behroozi13b}.

It is unclear whether the observed pace of galaxy size evolution
($\reff\propto H(z)^{-2/3}$) implies that $\reff$ evolves in
proportion to $R_{\rm{halo}}$, as may be expected in the case that
galaxy size scales with halo size in the present-day universe
\citep{kravtsov13}. It may be a coincidence that the observed pace of
evolution is the same as that for halo radii with respect to the
critical density, and it appears more natural to expect galaxy sizes
to scale with halo mass and radius that are defined in terms of matter
or virial density.  In this spirit, the tendency for late-type
galaxies to display rather slower size evolution than expected has
been given ample attention in the literature.

\citet{somerville08} argued that because halos are less concentrated
at high redshift, baryonic disks are larger in proportion to the
virial radii of halos, leading to slower size evolution.  In addition,
\citet{dutton11} showed that accreting, gaseous disks with a simple
but self-consistent prescription for star formation lead to similarly
slow evolution of the stellar disk scale radius as a result of
recycling of gas and radial variations in star formation.  In
addition, stellar feedback may have a more direct impact on disk sizes
as low-angular momentum material is ejected \citep[e.g.,][]{maller02,
  brook11}.

The sizes we measured are the not strictly disk scale lengths, as we
sample the whole galaxy, including the bulge.  Therefore, bulge
formation in late-type galaxies slows down size evolution as
parameterized here.  Bulge formation can either occur rapidly, through
mergers \citep{toomre72} or clump formation and migration in unstable
disks \citep{dekel09,ceverino10}, or gradually, through secular
evolution driven by non-axisymmetries in the disk potential
\citep{kormendy04}.  The prediction of any of these scenarios is that
the galaxies with higher global S\'ersic indices will have smaller
sizes at a given mass.  The observation that evolution is faster at
$z>2$ and slower at $z<1$ (see Section \ref{sec:complate} and Figure
\ref{lbg}), combined with the appearance of redder, more compact
galaxies at late cosmic times, suggests that bulge formation plays an
important role in the evolution of half-light radii of late-type
galaxies.

\subsection{Evolution of Massive Early-type Galaxies}\label{sec:early}

The co-moving number density of $L^*$ early types has strongly
increased over the redshift range examined in this paper ($0<z<3$), as
was shown by, e.g., \citet{bell04, faber07, brown07, ilbert10,
  brammer11, buitrago13, muzzin13}.  Here this is illustrated in
Figures \ref{rhist} and \ref{rc}.  Naturally, the progenitors of the
newly formed early-type galaxies must be looked for among the
star-forming, late-type population.  The skewed size distribution of
late types toward small sizes (see Section \ref{sec:skew} and Figure
\ref{rhist}) points at the existence of a population of small
late-type galaxies that span the entire size range seen for early-type
galaxies.  Figure \ref{rc} illustrates that this is the case over
essentially the entire redshift range probed by our sample.

The tail of small star-forming galaxies shown in Figure \ref{rc} at
$z>1.5$ \citep[also see][]{barro13, williams14, barro14} may reflect
the intriguing possibility of a scenario in which such small yet
massive star-forming galaxies are the immediate progenitors of compact
early-type galaxies.  Their number density does not rapidly change
between $z=3$ and $z=1.5$, whereas the number of early-type galaxies
does rapidly increase over that redshift range (see Figure \ref{rc}).
This would suggest the continuous emergence of additional small
late-type galaxies that represent a transitional phase between the
bulk of the late-type population and the early-type population, as
recently advocated by \citet{dekel14} on the basis of analytical
arguments and simulations.

An alternative interpretation is that the star-forming population
consists of ``normal'' late types and a population of early-type
galaxies that revived their star-formation activity.  The simplest
implementation of this model, in which these ``frosting'' early types
have the same size distribution as the quiescent early types, can be
ruled out: the skewed size distribution of late types is not well
described by two log-normal distributions centered at the respective
peaks of the size distributions for late- and early-type galaxies.  In
general, the size distribution of the full population of galaxies
(early and late types combined) is not observed to be bimodal in the
sense that there is no clear gap between two fiducial populations of
small and large galaxies, nor can the size distribution be accurately
represented by a single Gaussian distribution.  More complicated
models of the ``frosting'' flavor, in which a large, star-forming disk
reassembles to surround a compact, quiescent component, cannot be
immediately ruled out.  However, such scenarios seem implausible as
the implied color and mass-to-light ratio gradients of such galaxies
would likely be stronger than observed \citep{wuyts12}.  Measurements
of the stellar density in the central regions of early- and late-type
galaxies can be used to provide further constraints.

Whether or not the small late-type galaxies represent a transitional
phase, the central idea in the formation of early-type galaxies is
that the formation of an early-type galaxy requires the formation of a
concentrated stellar body with a high density \citep[e.g.,][]{franx08,
  bell12}.  One possibility is that a substantial amount of material
flows to the center under the influence of mergers
\citep[e.g.,][]{dimatteo05} or violently unstable disks and clump
formation/migration \citep{dekel09,ceverino10,dekel14}.  It remains to
be seen whether such processes can reproduce the correct stellar
density profiles \citep{wuyts10}.

As we showed in Section \ref{sec:compearly} and Figures \ref{rc} and
\ref{small}, the number density of small, compact early-type galaxies
strongly decreases between $z\sim 1.5$ and the present.  This
immediately implies that early-type galaxies, after they first form as
compact, quiescent objects, have to substantially grow in size over
time.  Combining this with the suggestion that new early-type galaxies
likely form out of the smallest late-type galaxies, the implication is
that early-type galaxies are the most dense \citep[and disk-like in
structure, e.g.,][]{vanderwel11a, bruce12, chang13} immediately after
their star formation is truncated.  The amount of later evolution in
size and density is dictated by the (non-evolving) slope of the
size-mass relation and the evolution of its intercept.  This naturally
fits into the general idea that a gas-rich formation phase is followed
by a more quiescent, dissipationless formation phase.

The scatter in the size-mass relation of $\sim 0.15-0.20$~dex (see
Figure \ref{parevol}) shows that there is some variation in the amount
of dissipative and dissipationless formation, yet, the fact that we
see little or no evolution in the size scatter, as predicted by
\citet{shankar13}, implies that the amount of dissipation integrated
over cosmic history does not vary greatly. Some early-type galaxies
may have experienced an intensely dissipative phase at early times,
while other -- similarly massive -- galaxies may have continued a less
intense star-forming phase up until recently.  The compact $z>1.5$
early-type galaxies would fall in the former category; the large,
massive star-forming galaxies at $z\sim 0.5-1.5$ may be the
progenitors of galaxies in the latter.

Within this framework, independent evidence for the increase in
stellar mass of individual early-type galaxies by a factor of 2 to 3
between $z=2$ and the present \citep{vandokkum10} implies that the
growth in size depends on the growth in mass as $\Delta \reff\propto
\Delta M_*^{\sim 2}$.  This steep dependence is consistent with a
merger scenario.

Satellite galaxies can be stripped and their stars deposited on
large-radius orbits.  Direct and stringent constraints on the minor
merger rate are difficult to obtain, but it has proved to be difficult
to observationally confirm a sufficiently large minor merger rate to
explain the observed evolution \citep{newman12}.  Mergers among
galaxies that occupy the size-mass relation for early-type galaxies,
that is, pure dry mergers, may not occur at sufficient rates
\citep[e.g.,][]{nipoti12}.

Alternatively, mergers between similarly massive galaxies with
different sizes can induce large changes in the size-mass distribution
of the population.  Assuming that the size distribution of progenitors
partaking in major mergers is the same as that of the population as a
whole, a $\reff=1$~kpc early-type galaxy at $z\sim 2$ will merge, on
average, with a late-type galaxy that is 3 times larger.  The
properties of the merger remnant will depend on the amount of
dissipation and the dynamics of the merger, but it is conceivable that
the remnant will be much larger than the compact progenitor.  A dense
inner region will remain in place, and the strong correlation between
central density and quiescence implies that the remnant is likely to
be quiescent as well.

The mass ratio distribution in the merger history of early-type
galaxies, and its effect on size evolution, will remain a topic of
debate.  However, merging can account for, and is arguably required to
explain, the disappearance from $z\sim 2$ to the present of disklike
structures among $L^*$ early types
\citep{vanderwel11a,bruce12,chang13}, and the observation that the
most massive galaxies in the present-day universe do not have a
disklike structure, but are intrinsically round \citep{vanderwel09b}.
A combined analysis of the evolution of size and morphological
properties \citep[see, e.g.,][]{huertas13} will aid to simultaneously
interpret size growth and disk destruction.

The above narrative shows that we have gathered a plausible set of
mechanisms that may play a role in explaining the formation and
subsequent evolution of early-type galaxies.  Despite this, we lack
the basis of a simple analytical framework that is similar to our
model for disk formation.  However, we note that the rapid pace of
size evolution is very close to the size evolution expected for halos
as defined by their virial density: $R\propto H(z)^{-1.24}$ for halos
compares well with $\reff\propto H(z)^{-1.29}$ for massive early-type
galaxies.  If we assume that these galaxies only grow through the
accretion of other halos and their stellar content, then it is perhaps
not a coincidence that halos and galaxies both follow the evolutionary
path expected for a dissipationless, top-hat collapse scenario.

\subsection{Evolution of Low-mass Early-type Galaxies}\label{sec:lowearly}

As we noted in Section \ref{sec:medsize} the slope of the size-mass
relation for early-type galaxies flattens below stellar mass $\sim
\msolb$, and the size evolution is more comparable to that of late
types than that of early types (see Figure \ref{zr}).  This suggests
that there is a population of low-mass early types that may have
formed out of late-type galaxies without going through a transitional
phase in which high central densities are attained.  The stripping of
gas from satellite galaxies is a natural explanation for such
evolution and can explain the existence of an excess population of
early-type galaxies in clusters that have structural properties
similar to those of late-type galaxies in the field
\citep{vanderwel10}. Satellites are common in this mass range in the
present-day universe \citep[e.g.,][]{vandenbosch08}, but not at higher
redshifts, lending the stripping scenario more credence on the basis
of based the rapid increase in the comoving number density since
$z\sim 1-1.5$ (see Figures \ref{mr} and \ref{rhist}).

On the other hand, the early types with mass $\lesssim \msolb$ are
$\sim 2$ times smaller than equally massive late types.  Disk fading
may contribute to this difference, but bulge formation and, in
general, processes that cause more massive galaxies to transform into
early-type galaxies, may play a role in the low-mass regime as well.
A model such as that presented by \citet{peng10} can be expanded in
order to separately reproduce the size-mass relations for different
types of ``quenched'' galaxies.

\section{Summary}\label{sec:summary}

In this paper we present the size-mass distribution of 30,958 galaxies
over a large range in mass ($>\msolc$) and redshift ($0<z<3$),
distinguishing between early-type and late-type galaxies on the basis
of their star-formation activity.  Spectroscopic and photometric
redshifts, stellar masses, and rest-frame properties are determined by
using data from the 3D-HST survey and auxiliary, multiwavelength
photometric data sets spanning from the $U$ band to 8~$\mu$m (see
Section \ref{sec:data}).  Galaxy sizes are measured from CANDELS
imaging by single-component S\'ersic profile fits to two-dimensional
light distributions, with a correction for (redshift-dependent) color
gradients (Section \ref{sec:size}).

Consistent with previous results, we find that high-redshift galaxies
are substantially smaller than equally massive, present-day
counterparts.  As is shown in Figures \ref{mr}, \ref{parevol} and
\ref{zr}, late-type galaxies are, on average, a factor of $\sim 2$
smaller at $z=2$ than at the present day, whereas for massive
early-type galaxies this is a factor of $\sim 4$.  We find that the
size evolution of late-type galaxies is marginally better described as
a function of the redshift-dependent Hubble parameter, $H(z)$, than as
a function of the scale factor, $1+z$ (Figure \ref{zr_res}).  Average
mass-matched sizes of late- and early-type galaxies evolve as $\reff
\propto H(z)^{-0.66} \propto (1+z)^{-.75}$ and $\reff\propto
H(z)^{-1.29} \propto (1+z)^{-1.48}$, respectively (Figure
\ref{parevol} and Table \ref{tab:fitsize}).

High-mass late-type galaxies evolve marginally faster than low-mass
late-type galaxies (Figure \ref{zr} and Table \ref{tab:medsize}), but
the data are consistent with no evolution in the overall slope of the
size-mass relation.  At all redshifts $z\le 3$ we find that the slope
is shallow for late-type galaxies ($\reff\propto M_*^{0.22}$ for
galaxies with stellar mass $M_*>3\times 10^9~\msol$) and is steep for
early-type galaxies ($\reff\propto M_*^{0.75}$ for galaxies with
stellar mass $M_*>2\times 10^{10}~\msol$).  The size-mass relation for
lower-mass early-type galaxies is more similar to that of late types
than that of high-mass early types (Section \ref{sec:lowearly}).  Once
cross-contamination between the two classes of galaxies and outliers
are taken into account (Figure \ref{parevol} and Section
\ref{sec:fitsize}), we also find no evidence for evolution in the
(intrinsic) size scatter at a fixed galaxy mass.  The implications of
these results are discussed in Section \ref{sec:discussion}.

The data presented here are consistent with essentially most published
data sets (Section \ref{sec:comparison}).  Because of the sample size
and dynamic range in mass and redshift, the immediate implications of
the measurements are less ambiguous than was the case for previous
studies.  In particular, we show in Figure \ref{rc} that the size
distribution of $z\sim 2$ early-type galaxies is significantly
different from that of any subset of low-redshift galaxies with the
same comoving number density; small early-type galaxies, which are
typical at $z\sim 2$, do not exist in equal numbers today (Figure
\ref{small}) and must therefore undergo strong size evolution in the
intervening time.

The size-mass distributions from the 3D-HST and CANDELS projects
presented here provide a solid framework for galaxy evolution models,
and strongly constrain the interplay between structure formation and
galaxy formation \citep[e.g.,][]{stringer13}.  The steadily evolving
intercept of the size-mass relation, in combination with the
non-evolving slope and scatter, present tight constraints on how
baryons condense and form galaxies at the centers of dark matter halos
(e.g., Section \ref{sec:late}).  The different assembly mechanisms of
early- and late-type galaxies act similarly at all redshifts, as
evidenced by the very different, but unchanging slopes of their
respective size-mass relations.

\acknowledgements{This work is based on observations taken by the
  CANDELS Multi-Cycle Treasury Program (PI: Faber) and the 3D-HST
  Treasury Program (PI: van Dokkum) with the NASA/ESA \emph{HST},
  which is operated by the Association of Universities for Research in
  Astronomy, Inc., under NASA contract NAS5-26555.}

\bibliographystyle{apj}

\begin{appendix}
  \section{Complementary Size Distributions}

  Throughout this paper we use the radius $\reff$ as measured along
  the major axis.  Circularized sizes ($R_{\rm{eff,circ}} = \reff
  \sqrt{b/a}$, where $b/a$ is the projected axis ratio) have often
  been used in the literature.  For this reason we provide the
  circularized size distributions for the early- and late-type samples
  in Table \ref{tab:medsize_circ}.  We stress that since galaxies are
  predominantly oblate, that is, disklike, using $\reff$ instead of
  $R_{\rm{eff,circ}}$ is more prudent: $\reff$ is (almost) independent
  of inclination, while $R_{\rm{eff,circ}}$ depends on the short
  projected axis, which obviously strongly varies with inclination.

  Throughout the paper we distinguish between late- and early-type
  galaxies on the basis of star formation activity.  For some purposes
  it may be more useful to work with the size distributions of the
  full sample, without separating by type.  In Table
  \ref{tab:medsize_all} we provide the size distributions of the full
  sample.

  Finally, since stellar mass is a derived model-dependent quantity
  that is potentially suffering from large systematic errors, one
  might be interested in galaxy size as a function of luminosity,
  which is essentially a directly observed quantity.  In Table
  \ref{tab:medsize_all} and \ref{tab:medsize_l} we provide the size
  distributions as a function of rest-frame $V$-band luminosity.

\begin{table*}
\begin{scriptsize}
\begin{center}
  \caption{Logarithmic Size Distribution (16\% -- 84\% range) as a
    Function of Galaxy Mass and Redshift.  Identical to Table
    \ref{tab:medsize}, but with circularized sizes instead of
    semi-major axis sizes.}
 \begin{tabular}{|cc||c|c|c|c|c||c|c|c|c|c|}
\hline
& & \multicolumn{5}{c||}{Early-Type Galaxies} & \multicolumn{5}{c|}{Late-Type Galaxies}\\
\cline{3-12}
 $z$ &  &  {\tiny $M_*=10^{9.25}$}  & 9.75 & 10.25 & 10.75 & 11.25 & 9.25 & 9.75 & 10.25 & 10.75 & 11.25 \\
\hline
     & 16\% & -0.07$\pm$0.08 & -0.05$\pm$0.04 & 0.02$\pm$0.03 & 0.29$\pm$0.04 & \nodata & 0.07$\pm$0.01 & 0.21$\pm$0.02 & 0.27$\pm$0.02 & 0.51$\pm$0.03 & \nodata \\
0.25 & 50\% & 0.17$\pm$0.01 & 0.17$\pm$0.03 & 0.24$\pm$0.04 & 0.57$\pm$0.03 & 0.72$\pm$0.07 & 0.30$\pm$0.01 & 0.42$\pm$0.01 & 0.49$\pm$0.03 & 0.70$\pm$0.02 & 0.91$\pm$0.04 \\
     & 84\% & 0.31$\pm$0.03 & 0.39$\pm$0.05 & 0.48$\pm$0.04 & 0.84$\pm$0.05 & 1.01$\pm$0.11 & 0.54$\pm$0.01 & 0.62$\pm$0.02 & 0.73$\pm$0.02 & 0.86$\pm$0.05 & 1.06$\pm$0.06 \\
\hline
     & 16\% & -0.11$\pm$0.04 & -0.25$\pm$0.02 & -0.11$\pm$0.02 & 0.11$\pm$0.02 & 0.57$\pm$0.02 & 0.04$\pm$0.01 & 0.17$\pm$0.01 & 0.27$\pm$0.01 & 0.39$\pm$0.01 & 0.64$\pm$0.04 \\
0.75 & 50\% & 0.12$\pm$0.02 & 0.10$\pm$0.02 & 0.13$\pm$0.02 & 0.36$\pm$0.01 & 0.74$\pm$0.03 & 0.26$\pm$0.00 & 0.39$\pm$0.01 & 0.49$\pm$0.01 & 0.61$\pm$0.02 & 0.82$\pm$0.03 \\
     & 84\% & 0.34$\pm$0.02 & 0.34$\pm$0.03 & 0.33$\pm$0.01 & 0.56$\pm$0.01 & 0.93$\pm$0.04 & 0.47$\pm$0.01 & 0.58$\pm$0.01 & 0.66$\pm$0.01 & 0.78$\pm$0.01 & 0.98$\pm$0.02 \\
\hline
     & 16\% & \nodata & -0.21$\pm$0.02 & -0.23$\pm$0.01 & -0.07$\pm$0.02 & 0.32$\pm$0.04 & -0.05$\pm$0.01 & 0.10$\pm$0.01 & 0.23$\pm$0.01 & 0.36$\pm$0.01 & 0.51$\pm$0.01 \\
1.25 & 50\% & \nodata & 0.06$\pm$0.03 & -0.01$\pm$0.02 & 0.16$\pm$0.02 & 0.49$\pm$0.03 & 0.19$\pm$0.00 & 0.32$\pm$0.00 & 0.42$\pm$0.00 & 0.53$\pm$0.01 & 0.67$\pm$0.02 \\
     & 84\% & \nodata & 0.33$\pm$0.03 & 0.27$\pm$0.03 & 0.44$\pm$0.03 & 0.75$\pm$0.03 & 0.41$\pm$0.01 & 0.51$\pm$0.01 & 0.60$\pm$0.01 & 0.69$\pm$0.01 & 0.83$\pm$0.01 \\
\hline
     & 16\% & \nodata & -0.12$\pm$0.06 & -0.36$\pm$0.02 & -0.14$\pm$0.02 & 0.11$\pm$0.06 & -0.09$\pm$0.01 & 0.02$\pm$0.01 & 0.16$\pm$0.01 & 0.24$\pm$0.02 & 0.41$\pm$0.03 \\
1.75 & 50\% & \nodata & 0.12$\pm$0.04 & -0.05$\pm$0.04 & 0.08$\pm$0.02 & 0.37$\pm$0.03 & 0.15$\pm$0.00 & 0.24$\pm$0.01 & 0.38$\pm$0.01 & 0.47$\pm$0.01 & 0.58$\pm$0.02 \\
     & 84\% & \nodata & 0.36$\pm$0.04 & 0.27$\pm$0.04 & 0.42$\pm$0.03 & 0.67$\pm$0.03 & 0.36$\pm$0.01 & 0.46$\pm$0.01 & 0.55$\pm$0.01 & 0.64$\pm$0.01 & 0.75$\pm$0.02 \\
\hline
     & 16\% & \nodata & \nodata & -0.53$\pm$0.09 & -0.34$\pm$0.03 & 0.02$\pm$0.05 & \nodata & -0.06$\pm$0.01 & 0.06$\pm$0.02 & 0.16$\pm$0.02 & 0.29$\pm$0.04 \\
2.25 & 50\% & \nodata & \nodata & -0.12$\pm$0.05 & -0.02$\pm$0.03 & 0.25$\pm$0.05 & \nodata & 0.17$\pm$0.01 & 0.29$\pm$0.01 & 0.39$\pm$0.01 & 0.52$\pm$0.02 \\
     & 84\% & \nodata & \nodata & 0.25$\pm$0.03 & 0.41$\pm$0.06 & 0.72$\pm$0.12 & \nodata & 0.38$\pm$0.01 & 0.49$\pm$0.01 & 0.56$\pm$0.01 & 0.70$\pm$0.02 \\
\hline
     & 16\% & \nodata & \nodata & \nodata & -0.37$\pm$0.05 & -0.08$\pm$0.11 & \nodata & \nodata & 0.01$\pm$0.02 & 0.08$\pm$0.05 & 0.24$\pm$0.03 \\
2.75 & 50\% & \nodata & \nodata & \nodata & -0.01$\pm$0.04 & 0.29$\pm$0.09 & \nodata & \nodata & 0.27$\pm$0.01 & 0.34$\pm$0.01 & 0.43$\pm$0.02 \\
     & 84\% & \nodata & \nodata & \nodata & 0.40$\pm$0.06 & 0.62$\pm$0.10 & \nodata & \nodata & 0.48$\pm$0.02 & 0.54$\pm$0.02 & 0.63$\pm$0.06 \\
\hline
\end{tabular}
\label{tab:medsize_circ}
\end{center}
\end{scriptsize}
\end{table*}

\begin{table*}
\begin{scriptsize}
\begin{center}
  \caption{Logarithmic Size Distributions (16\% -- 84\% range) as a
    Function of Rest-frame $V$-band Luminosity and Redshift.}
 \begin{tabular}{|cc||c|c|c|c|c||c|c|c|c|c|}
\hline
& & \multicolumn{5}{c||}{Early-Type Galaxies} & \multicolumn{5}{c|}{Late-Type Galaxies}\\
\cline{3-12}
 $z$ &  &  {\tiny $L_*=10^{9.25}$}  & 9.75 & 10.25 & 10.75 & 11.25 & 9.25 & 9.75 & 10.25 & 10.75 & 11.25 \\
\hline
     & 16\% & 0.04$\pm$0.05 & 0.06$\pm$0.03 & 0.21$\pm$0.05 & 0.61$\pm$0.07 & \nodata & 0.16$\pm$0.03 & 0.31$\pm$0.02 & 0.42$\pm$0.02 & 0.68$\pm$0.06 & \nodata \\
0.25 & 50\% & 0.27$\pm$0.02 & 0.28$\pm$0.02 & 0.45$\pm$0.02 & 0.72$\pm$0.03 & \nodata & 0.42$\pm$0.02 & 0.57$\pm$0.01 & 0.69$\pm$0.02 & 0.92$\pm$0.03 & \nodata \\
     & 84\% & 0.44$\pm$0.04 & 0.51$\pm$0.04 & 0.72$\pm$0.05 & 0.99$\pm$0.06 & \nodata & 0.64$\pm$0.02 & 0.76$\pm$0.01 & 0.87$\pm$0.02 & 1.07$\pm$0.04 & \nodata \\
\hline
     & 16\% & -0.01$\pm$0.05 & -0.08$\pm$0.03 & 0.05$\pm$0.02 & 0.31$\pm$0.01 & 0.54$\pm$0.05 & 0.06$\pm$0.02 & 0.21$\pm$0.01 & 0.37$\pm$0.01 & 0.54$\pm$0.01 & 0.73$\pm$0.10 \\
0.75 & 50\% & 0.24$\pm$0.02 & 0.21$\pm$0.02 & 0.27$\pm$0.02 & 0.54$\pm$0.02 & 0.81$\pm$0.06 & 0.33$\pm$0.01 & 0.46$\pm$0.01 & 0.60$\pm$0.01 & 0.75$\pm$0.01 & 0.93$\pm$0.08 \\
     & 84\% & 0.41$\pm$0.02 & 0.46$\pm$0.02 & 0.47$\pm$0.02 & 0.81$\pm$0.02 & 1.03$\pm$0.06 & 0.57$\pm$0.02 & 0.67$\pm$0.00 & 0.79$\pm$0.01 & 0.93$\pm$0.02 & 1.12$\pm$0.08 \\
\hline
     & 16\% & -0.10$\pm$0.12 & -0.12$\pm$0.04 & -0.14$\pm$0.02 & 0.07$\pm$0.02 & 0.29$\pm$0.03 & 0.03$\pm$0.02 & 0.11$\pm$0.01 & 0.25$\pm$0.01 & 0.43$\pm$0.01 & 0.48$\pm$0.05 \\
1.25 & 50\% & 0.19$\pm$0.04 & 0.23$\pm$0.03 & 0.10$\pm$0.01 & 0.32$\pm$0.01 & 0.56$\pm$0.03 & 0.32$\pm$0.01 & 0.37$\pm$0.01 & 0.49$\pm$0.01 & 0.64$\pm$0.01 & 0.75$\pm$0.02 \\
     & 84\% & 0.45$\pm$0.11 & 0.51$\pm$0.03 & 0.40$\pm$0.02 & 0.58$\pm$0.03 & 0.83$\pm$0.04 & 0.66$\pm$0.04 & 0.60$\pm$0.01 & 0.70$\pm$0.00 & 0.82$\pm$0.01 & 0.96$\pm$0.03 \\
\hline
     & 16\% & \nodata & -0.04$\pm$0.08 & -0.19$\pm$0.03 & -0.11$\pm$0.03 & 0.20$\pm$0.03 & 0.12$\pm$0.03 & 0.05$\pm$0.01 & 0.15$\pm$0.01 & 0.32$\pm$0.01 & 0.37$\pm$0.08 \\
1.75 & 50\% & \nodata & 0.27$\pm$0.03 & 0.09$\pm$0.02 & 0.15$\pm$0.02 & 0.37$\pm$0.02 & 0.43$\pm$0.04 & 0.32$\pm$0.01 & 0.40$\pm$0.01 & 0.56$\pm$0.01 & 0.70$\pm$0.03 \\
     & 84\% & \nodata & 0.53$\pm$0.05 & 0.39$\pm$0.03 & 0.50$\pm$0.04 & 0.71$\pm$0.04 & 0.82$\pm$0.05 & 0.58$\pm$0.01 & 0.62$\pm$0.01 & 0.75$\pm$0.01 & 0.87$\pm$0.01 \\
\hline
     & 16\% & \nodata & \nodata & -0.26$\pm$0.11 & -0.26$\pm$0.03 & 0.00$\pm$0.05 & \nodata  & 0.01$\pm$0.04 & 0.08$\pm$0.01 & 0.20$\pm$0.01 & 0.26$\pm$0.06 \\
2.25 & 50\% & \nodata & \nodata & 0.08$\pm$0.04 & 0.07$\pm$0.04 & 0.24$\pm$0.03 & \nodata & 0.30$\pm$0.03 & 0.34$\pm$0.01 & 0.44$\pm$0.01 & 0.59$\pm$0.02 \\
     & 84\% & \nodata & \nodata & 0.35$\pm$0.08 & 0.50$\pm$0.05 & 0.49$\pm$0.05& \nodata & 0.68$\pm$0.06 & 0.56$\pm$0.01 & 0.65$\pm$0.01 & 0.79$\pm$0.03 \\
\hline
     & 16\% & \nodata & \nodata & \nodata & -0.27$\pm$0.03 & 0.01$\pm$0.05 & \nodata & \nodata & 0.09$\pm$0.01 & 0.18$\pm$0.01 & 0.26$\pm$0.07 \\
2.75 & 50\% & \nodata & \nodata & \nodata & 0.03$\pm$0.05 & 0.34$\pm$0.09 & \nodata & \nodata & 0.33$\pm$0.02 & 0.43$\pm$0.01 & 0.54$\pm$0.02 \\
     & 84\% & \nodata & \nodata & \nodata & 0.42$\pm$0.07 & 0.73$\pm$0.08 & \nodata & \nodata & 0.62$\pm$0.04 & 0.63$\pm$0.01 & 0.75$\pm$0.02 \\
\hline
\end{tabular}
\label{tab:medsize_l}
\end{center}
\end{scriptsize}
\end{table*}

\begin{table*}
\begin{scriptsize}
\begin{center}
  \caption{Logarithmic size distributions (16\% -- 84\% range)
    for the full population (early- and late-type galaxies combined)
    as a function of mass and redshift, and rest-frame $V$-band
    luminosity and redshift.}
 \begin{tabular}{|cc||c|c|c|c|c||c|c|c|c|c|}
\hline
& & \multicolumn{10}{c|}{Early+Late Type Galaxies}\\
\cline{3-12}
 $z$ &  &  {\tiny $M_*=10^{9.25}$}  & 9.75 & 10.25 & 10.75 & 11.25 & {\tiny $L_*=10^{9.25}$} & 9.75 & 10.25 & 10.75 & 11.25 \\
\hline
     & 16\% & 0.21$\pm$0.01 & 0.28$\pm$0.02 & 0.27$\pm$0.03 & 0.49$\pm$0.04 & \nodata 	& 0.14$\pm$0.02 & 0.26$\pm$0.02 & 0.37$\pm$0.01 & 0.62$\pm$0.04 & \nodata \\
0.25 & 50\% & 0.46$\pm$0.01 & 0.54$\pm$0.02 & 0.54$\pm$0.03 & 0.75$\pm$0.03 & \nodata 	& 0.37$\pm$0.02 & 0.53$\pm$0.01 & 0.62$\pm$0.02 & 0.79$\pm$0.03 & \nodata \\
     & 84\% & 0.69$\pm$0.01 & 0.77$\pm$0.02 & 0.82$\pm$0.02 & 0.99$\pm$0.02 & \nodata 	& 0.62$\pm$0.02 & 0.74$\pm$0.01 & 0.85$\pm$0.01 & 1.05$\pm$0.03 & \nodata \\
\hline
     & 16\% & 0.16$\pm$0.01 & 0.25$\pm$0.01 & 0.18$\pm$0.01 & 0.35$\pm$0.02 & 0.66$\pm$0.02 	& 0.05$\pm$0.02 & 0.18$\pm$0.01 & 0.28$\pm$0.01 & 0.40$\pm$0.02 & 0.62$\pm$0.06 \\
0.75 & 50\% & 0.41$\pm$0.01 & 0.52$\pm$0.01 & 0.52$\pm$0.01 & 0.59$\pm$0.01 & 0.85$\pm$0.01 	& 0.29$\pm$0.01 & 0.44$\pm$0.01 & 0.54$\pm$0.01 & 0.68$\pm$0.01 & 0.83$\pm$0.04 \\
     & 84\% & 0.64$\pm$0.01 & 0.74$\pm$0.01 & 0.78$\pm$0.01 & 0.84$\pm$0.01 & 1.04$\pm$0.02 	& 0.54$\pm$0.02 & 0.66$\pm$0.00 & 0.76$\pm$0.01 & 0.90$\pm$0.01 & 1.04$\pm$0.04 \\
\hline
     & 16\% & \nodata & 0.21$\pm$0.01 & 0.15$\pm$0.02 & 0.23$\pm$0.02 & 0.49$\pm$0.03 	& 0.02$\pm$0.02 & 0.10$\pm$0.01 & 0.19$\pm$0.01 & 0.29$\pm$0.01 & 0.36$\pm$0.04 \\
1.25 & 50\% & \nodata & 0.47$\pm$0.01 & 0.52$\pm$0.01 & 0.57$\pm$0.01 & 0.74$\pm$0.02 	& 0.30$\pm$0.02 & 0.36$\pm$0.01 & 0.47$\pm$0.01 & 0.58$\pm$0.01 & 0.67$\pm$0.04 \\
     & 84\% & \nodata & 0.68$\pm$0.01 & 0.74$\pm$0.01 & 0.80$\pm$0.01 & 0.94$\pm$0.02 	& 0.64$\pm$0.03 & 0.60$\pm$0.01 & 0.69$\pm$0.01 & 0.79$\pm$0.01 & 0.91$\pm$0.03 \\
\hline
     & 16\% & \nodata & 0.15$\pm$0.01 & 0.12$\pm$0.02 & 0.09$\pm$0.01 & 0.34$\pm$0.02 	& \nodata & 0.05$\pm$0.01 & 0.12$\pm$0.01 & 0.16$\pm$0.01 & 0.22$\pm$0.02 \\
1.75 & 50\% & \nodata & 0.42$\pm$0.01 & 0.48$\pm$0.01 & 0.48$\pm$0.01 & 0.64$\pm$0.03 	& \nodata & 0.32$\pm$0.01 & 0.39$\pm$0.01 & 0.51$\pm$0.01 & 0.51$\pm$0.04 \\
     & 84\% & \nodata & 0.65$\pm$0.01 & 0.69$\pm$0.01 & 0.74$\pm$0.01 & 0.83$\pm$0.02 	& \nodata & 0.58$\pm$0.01 & 0.61$\pm$0.00 & 0.73$\pm$0.01 & 0.83$\pm$0.02 \\
\hline
     & 16\% & \nodata & \nodata & 0.10$\pm$0.02 & 0.03$\pm$0.03 & 0.28$\pm$0.03 	& \nodata & \nodata & 0.07$\pm$0.01 & 0.15$\pm$0.01 & 0.10$\pm$0.03 \\
2.25 & 50\% & \nodata & \nodata & 0.41$\pm$0.01 & 0.45$\pm$0.02 & 0.59$\pm$0.03 	& \nodata & \nodata & 0.33$\pm$0.01 & 0.43$\pm$0.01 & 0.45$\pm$0.03 \\
     & 84\% & \nodata & \nodata & 0.63$\pm$0.01 & 0.68$\pm$0.01 & 0.83$\pm$0.02 	& \nodata & \nodata & 0.55$\pm$0.01 & 0.64$\pm$0.01 & 0.74$\pm$0.03 \\
\hline
     & 16\% & \nodata & \nodata & \nodata & 0.01$\pm$0.03 & 0.27$\pm$0.06 	& \nodata & \nodata & \nodata & 0.11$\pm$0.02 & 0.12$\pm$0.06 \\
2.75 & 50\% & \nodata & \nodata & \nodata & 0.43$\pm$0.01 & 0.52$\pm$0.03 	& \nodata & \nodata & \nodata & 0.41$\pm$0.01 & 0.52$\pm$0.03 \\
     & 84\% & \nodata & \nodata & \nodata & 0.65$\pm$0.02 & 0.75$\pm$0.04 	& \nodata & \nodata & \nodata & 0.62$\pm$0.01 & 0.74$\pm$0.02 \\
\hline
\end{tabular}
\label{tab:medsize_all}
\end{center}
\end{scriptsize}
\end{table*}

\end{appendix}

\end{document}